\documentclass[%
 reprint,
 amsmath,amssymb,
 aps,
]{revtex4-2}
\usepackage{graphicx,epsfig,units}
\usepackage{xcolor} 
\usepackage{soul} 
\usepackage{chemformula}
\usepackage{multirow}
\usepackage{booktabs}

\usepackage{tabularx}

\usepackage[flushleft]{threeparttable}
\usepackage[super]{nth}
\usepackage{float} 
\usepackage{amsmath,amsfonts,mathrsfs,amsbsy,bm,babel}
\usepackage{natbib}
\usepackage[caption=false]{subfig}
\usepackage{dblfloatfix} 
\usepackage{array,booktabs,arydshln,xcolor}

\usepackage{romannum}
\usepackage{graphicx}
\usepackage{comment}
\usepackage{blindtext}

\usepackage{hyperref}
\usepackage{appendix}
\hypersetup{
    colorlinks=true,
    linkcolor=blue,
    filecolor=magenta,      
    urlcolor=black,
}

\usepackage[normalem]{ulem}

\newcommand{\add}[1]{ { \color{blue}  #1 }}

\begin{document}
\pagenumbering{arabic}
\pagestyle{plain}
\title{General Solution for Elastic Networks on Arbitrary Curved Surfaces in the Absence of Rotational Symmetry}

\author{Yankang Liu}
\affiliation{Department of Physics and Astronomy, University of California Riverside, Riverside, California 92521, United States}

\author{Siyu Li}
\affiliation{Department of Physics and Astronomy, University of California Riverside, Riverside, California 92521, United States}

\author{Roya Zandi}
\affiliation{Department of Physics and Astronomy, University of California Riverside, Riverside, California 92521, United States}

\author{Alex~Travesset}
\affiliation{Department of Physics and Astronomy, Iowa State University and Ames Lab, Ames, Iowa 50011, United States}

\begin{abstract}

Understanding crystal growth over arbitrary curved surfaces with arbitrary boundaries is a formidable challenge, stemming from the complexity of formulating non-linear elasticity using geometric invariant quantities. Solutions are generally confined to systems exhibiting rotational symmetry. In this paper, we introduce a framework to address these challenges by numerically solving these equations without relying on inherent symmetries. We illustrate our approach by computing the minimum energy required for an elastic network containing a disclination at any point and by investigating surfaces that lack rotational symmetry. Our findings reveal that the transition from a defect-free structure to a stable state with a single 5-fold or 7-fold disclination strongly depends on the shape of the domain, emphasizing the profound influence of edge geometry. We discuss the implications of our results for general experimental systems, particularly in elucidating the assembly pathways of virus capsids. This research enhances our understanding of crystal growth on complex surfaces and expands its applications across diverse scientific domains.


\end{abstract}
\maketitle


The problem of crystal growth on curved geometries arises in many areas of science and technology. Understanding this phenomenon not only helps in the analysis of various materials—such as liquid crystals, curved arrays of microlenses in optical engineering systems, viral shells, encapsulins, carboxysomes, clathrin vesicles, and numerous other cellular organelles—but also provides a valuable paradigm for studying nonlinear elasticity.

As a crystal grows on curved surfaces, the location of defects (disclinations) is crucial in determining the topology and stability of the final structure. There have been many studies aimed at characterizing the defects (disclinations) arrangements necessary to minimize elastic energy\cite{Perez-GarridoDodgsonMoore1997, Bowick2000, Bowick2001, BowickMe2002, Schneider2005, Bowick2006, Bendito2007, GiomiBowick2007,Giomi2008,MorozovBruinsma2010, AzadiGrason2014, Meng2014, Azadi2016}. While we have a well-developed understanding of defect locations in crystals that grow with preserved rotational symmetry ~\cite{LiTravesset2019,Li2019,YinanMe2022}, our knowledge of defect formation on arbitrary surfaces with arbitrary boundary conditions is still limited. The long-standing problem of locating defects in a growing crystal in the absence of rotational symmetry is complex due to the conflict between maintaining equitriangular order and dealing with nonzero (positive or negative) Gaussian curvature.
\begin{figure}
    \centering
    \includegraphics[width=0.48\textwidth]{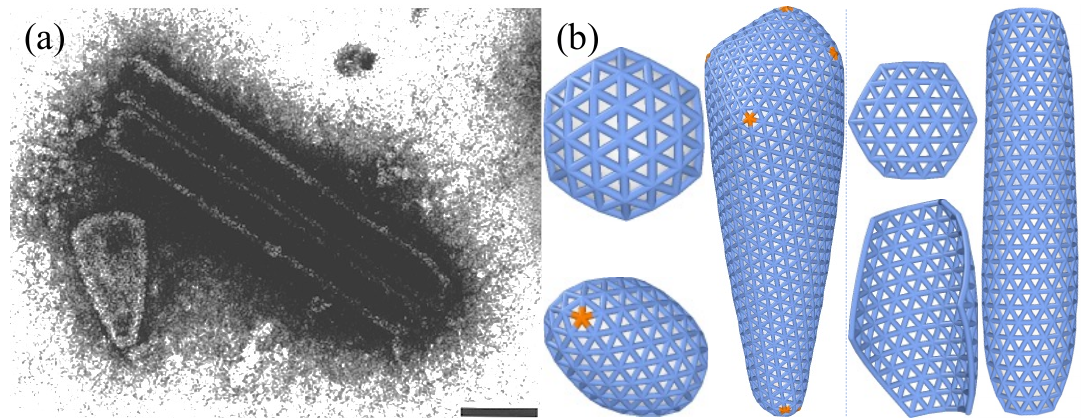}
    \caption{(a) Cryo-EM images of human immunodeficiency virus type 1 (HIV-1), adopted from Ref.~\cite{Ganser1999}. Both cylindrical and conical structures are observed in {\it in vitro} self-assembly studies. (b) Simulations of the formation of conical and cylindrical shells. At the top left, a hexagonal sheet initially grows without defects. However, due to the spontaneous curvature of the growing cap, pentameric defects appear, resulting in the formation of a conical structure. The emergence of the first disclination is highlighted in orange at the bottom left. On the right panel, defect-free growth results in a cylindrical structure.}
    \label{main:fig:cryoHIV}
\end{figure}
One compelling example of how the presence of disclinations (pentameric defects) can fundamentally alter the shell's topology is found in the conical structure of HIV capsids. Figure~\ref{main:fig:cryoHIV}a shows cryo-EM images from {\it in vitro} self-assembly studies of pure recombinant capsid proteins (CA) of HIV-1 \cite{Ganser1999}. Cylindrical and conical structures can coexist in the same solution; conical shells have 12 pentamers, while cylindrical shells have none.

Figure~\ref{main:fig:cryoHIV}b illustrates the simulations of the formation of conical and cylindrical structures \cite{nature2016}. The key difference between them lies in the formation of the first pentamer, which induces curvature and promotes the formation of additional pentamers. {In the absence of pentamers, the shell tends to form a cylindrical structure (see appendix for details)}. However, the formation of the first pentamer disrupts rotational symmetry in the structure's growth, resulting in a conical shape. 

The experiments and simulations in Fig.~\ref{main:fig:cryoHIV} confirm the role of disclination locations in determining the final structure.  Linear theories have long been used to predict the locations of disclinations. However, within linear theory, not only fundamental geometric principles, such as the Euler theorem limiting the number of disclinations to 12, may be violated~\cite{Castelnovo2017}, but significant errors also arise from ignoring the non-linear terms\cite{YinanMe2022}. {Previous work was constrained to one-dimensional systems due to restrictive assumptions: the disclination had to be either centrally located or absent, the boundary was limited to a circular shape, and the surface was required to exhibit rotational symmetry. These simplifications significantly limited the scope of the analysis. In contrast, the simulations presented in \cite{LiTravesset2018} and Fig.~\ref{main:fig:cryoHIV} operate without these constraints, revealing the critical need for a more comprehensive and robust solution to fully capture the complexity of the system.}

\begin{figure}
    \centering
    \includegraphics[width=\linewidth]{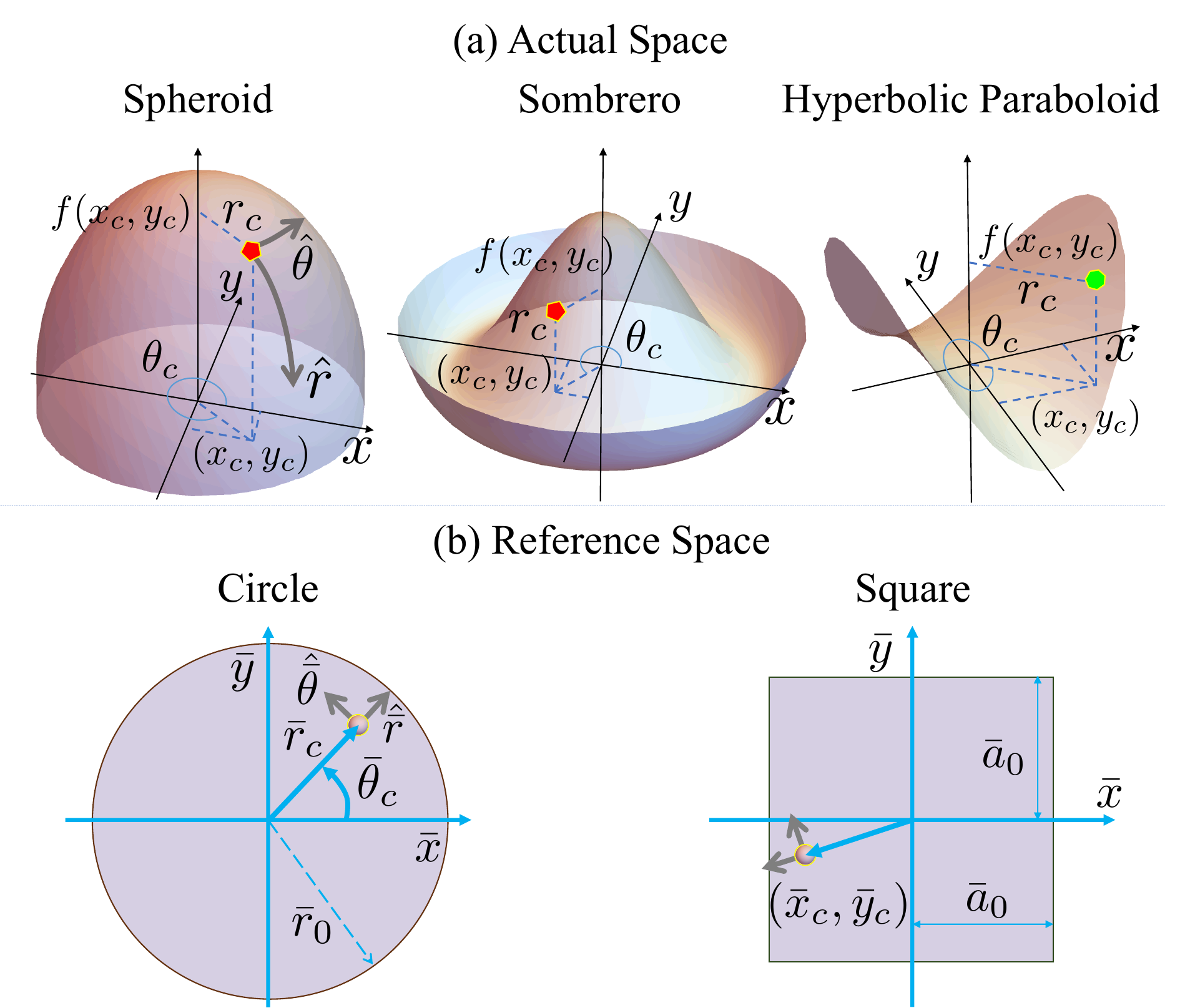}
    \caption{Illustration for actual and reference space. (a) The 5-fold disclination at position $(x_c,y_c,f(x_c,y_c))$ is denoted by a pentagon in red on a spheroid and a sombrero, and a 7-fold defect by a heptagon in green on a hyperbolic paraboloid. The projection of the disclinations in $x$ and $y$ plane are expressed in terms of cylindrical coordinates $(r_c,\theta_c)$. The disclination is able to move, indicated by the gray arrow on the spheroid. (b) The reference domain is defined such that $\bar{x}^2 + \bar{y}^2 \leq \bar{r}_0^2$, and $|\bar{x}|, |\bar{y}| \leq \bar{a}_0$. 
    }
    \label{main:fig:geometry}
\end{figure}



{In this paper, we present the general solution to these equations without imposing any symmetry constraints, introducing additional degrees of freedom. While this makes the growth process more realistic, it also increases the complexity of the system, making the resulting equations extremely difficult to solve. Notably, while solutions to these elasticity equations can be derived for any geometry, they must be obtained numerically, particularly in the presence of disclinations, due to their inherent complexity.} To illustrate the approach, we investigate the energetics of a single disclination on a domain of a curved surface, and explore the conditions under which the appearance of the first disclination becomes energetically favorable.  We consider non-revolutionary surfaces with both negative and positive curvatures and examine different choices of boundary geometries (see Fig.~\ref{main:fig:geometry}). {The generality of the method allows us to better understand the conditions under which the first disclination appears, enabling us to describe shell growth and the simulations more accurately.}


The fundamental aspect of our approach involves considering a {\em reference} space and an {\em actual} space.{Figure~\ref{main:fig:geometry} shows the surfaces defining the actual space alongside the domains in the reference space. It is important to note that both spaces are represented as the continuous limit of the lattice model \cite{Seung1988,NelsonBook2002}. From a discrete perspective, the reference space is depicted by a lattice composed of equilateral triangles, which may contain an arbitrary distribution of positive or negative disclinations.} In some situations, such as a buckled hexagonal sheet with a disclination (pentagon) at its center (see Fig.~S1), a cone or an icosahedron, the lattice in reference space can be visualized in actual space. However, for other cases, such as with a negative disclination, this visualization may not be feasible. 
To explore the geometric nature of the reference space when there exists one disclination $q=\pm 1$ at position $(\bar{x}_c,\bar{y}_c)$, the reference metric $\bar{g}_{\alpha\beta}(\bar{\mathbf{x}})$ is,
\begin{equation}\label{Eq:intro:reference_metric}
d\bar{s}^2 = d\bar{x}^2+d\bar{y}^2+(\alpha^2-1)\frac{\left[(\bar{x}-\bar{x}_c)d\bar{y}-(\bar{y}-\bar{y}_c)d\bar{x}\right]^2}{(\bar{x}-\bar{x}_c)^2+(\bar{y}-\bar{y}_c)^2}
\end{equation}
with $\alpha=1-q/6$ and $\bar{\mathbf{x}}=(\bar{x},\bar{y})$ the coordinate in the reference space. 
The Gaussian curvature of the reference metric is zero, except at $(\bar{x}_c,\bar{y}_c)$ \cite{LiTravesset2019}. If $a_L$ is the lattice constant in the reference space and the number of particles $N >> 1$, then the crystal area $\hat{A}$ {in the reference space} is
\begin{equation}\label{Eq:intro:area_reference}
    \hat{A}=\int d^2\bm{x} \sqrt{\bar{g}(\bm{x})}=\int d^2{\bar{\bm x}}\sqrt{\bar{g}(\bar{\bm x})}=\frac{\sqrt{3}}{4} N a_L^2 \ .
\end{equation}
We find $\sqrt{\bar{g}}=\alpha$ indicating that the area remains unchanged regardless of the location $(\bar{x}_c,\bar{y}_c)$ of the disclination. 

The actual space, on the other hand, represents the real surface where the lattice resides. The particles strive to approximate a triangular lattice as closely as possible in actual space, but due to {Gaussian curvature}, there are non-zero strains that incur an elastic energy cost. It is important to note that Gaussian curvature is quantized in units of $\pi/3$ in the reference space
, whereas it varies smoothly in the actual physical space. Consequently, the elastic energy is zero only in cases where Gaussian curvatures can be trivially mapped from the reference to actual space using an identity mapping.

{While our formalism can be applied to any surfaces,} for the purpose of this paper, we consider the actual space to consist of spheroids described by $f(r) = \beta \sqrt{R_0^2-r^2}$, sombreros $f(r) = {\beta R_0}/3(1-({r}/{R_0})^2+({r}/{R_0})^4)^{3/2}$,
or hyperbolic paraboloids $f(x,y) = x^2 / R_0^2 - \beta^2 y^2 / R_0^2$,
which 
{are not surfaces} of revolution and 
{have} a negative Gaussian curvature. Representative configurations of these three surfaces are shown in Fig.~\ref{main:fig:geometry}a. In all cases, $\beta$ is a dimensionless constant and $R_0$ defines the scales of the axes. The curvature increases as $\beta$ does. The metric in actual space is given in Eq.~\eqref{Eq:intro:actual_metric}.

To determine if a disclination is needed and, if so, to find its optimal position as a function of the growing crystal's area, we calculate the free energy of the curved surface \cite{YinanMe2022},
\begin{eqnarray}\label{Eq:intro:free_dens}
   F &=& F^{elastic} + F^{bending} + F^{abs}+F^{line} \\ \nonumber 
   &=&\int d^2{\bm x}\sqrt{g}\left[{\cal F}^{elastic} + {\cal F}^{bending}\right]+F^{abs}+F^{line},
\end{eqnarray}
where the first and second terms are the stretching and bending energies, the third term represents the attractive monomer-monomer interaction promoting crystal growth, and the last term is associated with the cost of the line tension due to the presence of a boundary.
The elastic term ${\cal F}^{elastic}$ (see Eq.~\eqref{Eq:app:elastic energy density}) contains a quadratic term \cite{LandauElasticityBook} in the strain tensor $u_{\alpha\beta}\equiv \left[g_{\alpha \beta}(\bm{x})-\bar{g}_{\alpha \beta}(\bm{\bar{x}})\right]/2$ 
(see Eqs.~\eqref{Eq:intro:reference_metric}~and~\eqref{Eq:intro:actual_metric}). 
The second term, in terms of the two radii of curvature $(R_i)_{i=1,2}$, is ${\cal F}^{bending} = \kappa[ (1/R_1-H_0)^2+ (1/R_2-H_0)^2]$
with $\kappa$ the bending rigidity and $H_0$ the mean spontaneous curvature of the subunits. The free energy density, Eq.~\eqref{Eq:intro:free_dens}, has no trivial solution. The only surfaces allowing zero strains have either zero Gaussian curvature: a plane, a cylinder ($q=0$) or {discrete} delta function of Gaussian curvatures, like a cone ($q=1$). Surfaces with vanishing bending rigidity have a constant curvature radius $R_1=R_2=1/H_0$, forming a sphere. There is, therefore, no surface that simultaneously minimizes both the elastic and bending energies. The third term in Eq.~\eqref{Eq:intro:free_dens}, $F^{abs}=-\Pi \hat{A} < 0$ with $\Pi$ the attractive interaction per unit area due to favorable hydrophobic contacts between subunits, is the driving force for crystal growth \cite{MorozovBruinsma2010} and the last term is the cost of the line tension due to the presence of a boundary.

For simplicity and to make the calculations more tractable, we consider a fixed geometry, where the actual metric $g_{\alpha\beta}({\bm x})$  is determined by the corresponding surface (see Appendix). Additionally, the defect distribution is given, so the reference metric $\bar{g}_{\alpha\beta}(\bar{\bm x})$ is also fixed. The problem then becomes solving for the elastic deformation between the reference space and the actual space. Specifically, we aim to obtain the mapping between the reference and the actual space,
\begin{equation}\label{Eq:transf_coordinates}
    {\bm x}={\cal U}(\bar{{\bm x}})\ \text{or}\ {\bar{\bm x}}={\cal U}_r({\bm x})\ .
\end{equation}
Following Ref.~\cite{LiTravesset2019,Efrati2009}, the minimization of the elastic energy given in Eq.~\eqref{Eq:intro:free_dens} leads to
\begin{equation}\label{Eq:intro:GM:covariant}
    \nabla_{\alpha} \sigma^{\alpha \beta}+(\bar{\Gamma}^{\beta}_{\gamma \nu}-{\Gamma}^{\beta}_{\gamma \nu})\sigma^{\gamma \nu} = 0 \ ,
\end{equation}
where 
$\sigma^{\alpha\beta}$ is the stress tensor and ${\bar{\Gamma}}^{\beta}_{\gamma \nu}$, ${\Gamma}^{\beta}_{\gamma \nu}$ are the Christoffel symbols for the reference and actual metrics, respectively, see Ref.~\cite{YinanMe2022}.

\begin{figure}
    \centering
    \includegraphics[width=\linewidth]{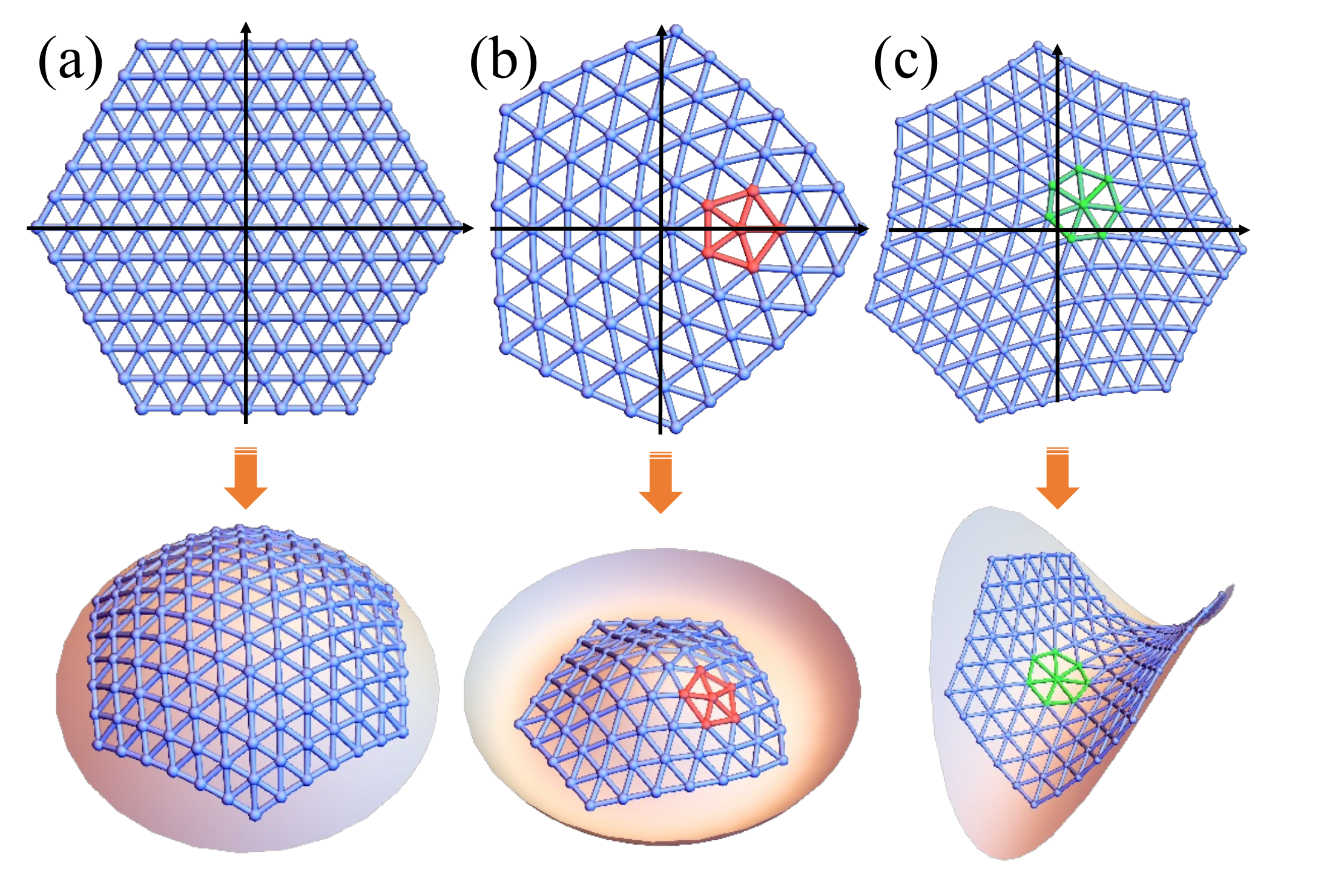}
    \caption{Elastic deformation visualized through lattice reconstruction. The first row shows triangular lattices constructed in the reference space and projected onto a plane. The second row maps these lattices, from left to right, to (a) a disclination-free lattice on a spheroid, (b) a lattice with an off-center 5-fold disclination on a sombrero, and (c) a lattice with an off-center 7-fold disclination on a hyperbolic paraboloid.}
    \label{main:fig:lattice:reconstruction}
\end{figure}
For simplicity, we solve Eq.~\eqref{Eq:intro:GM:covariant} under a tensionless boundary condition $\tau = 0$, and we fix {the Poisson ratio} $\nu_p=0.8$ unless stated otherwise, see SI~\Romannum{1}~and~\Romannum{2}. To uniquely determine the elastic deformation Eq.~\eqref{Eq:transf_coordinates}, we supplement an additional boundary condition: setting the origin at $\bar{x}, \bar{y} = 0$ to correspond to $x = 0, y = 0$. Then using finite element method~\cite{fenicsx}, we numerically solve the elastic equation~\eqref{Eq:intro:GM:covariant} to obtain the solution $\bm{x}(\bar{\bm{x}})$. This approach enables us to predict the position of any particle in actual space, given its location in the reference space, as illustrated in Fig.~\ref{main:fig:lattice:reconstruction}. The solution then, immediately gives the position of the disclination in the actual space once it is placed in the reference space.
Because particle positions in reference space are projected onto a plane for visualization, the triangles do not appear as equilateral, if a disclination is present (see the first row in Fig.~\ref{main:fig:lattice:reconstruction}).

\begin{figure}
    \centering
    \includegraphics[width=\linewidth]{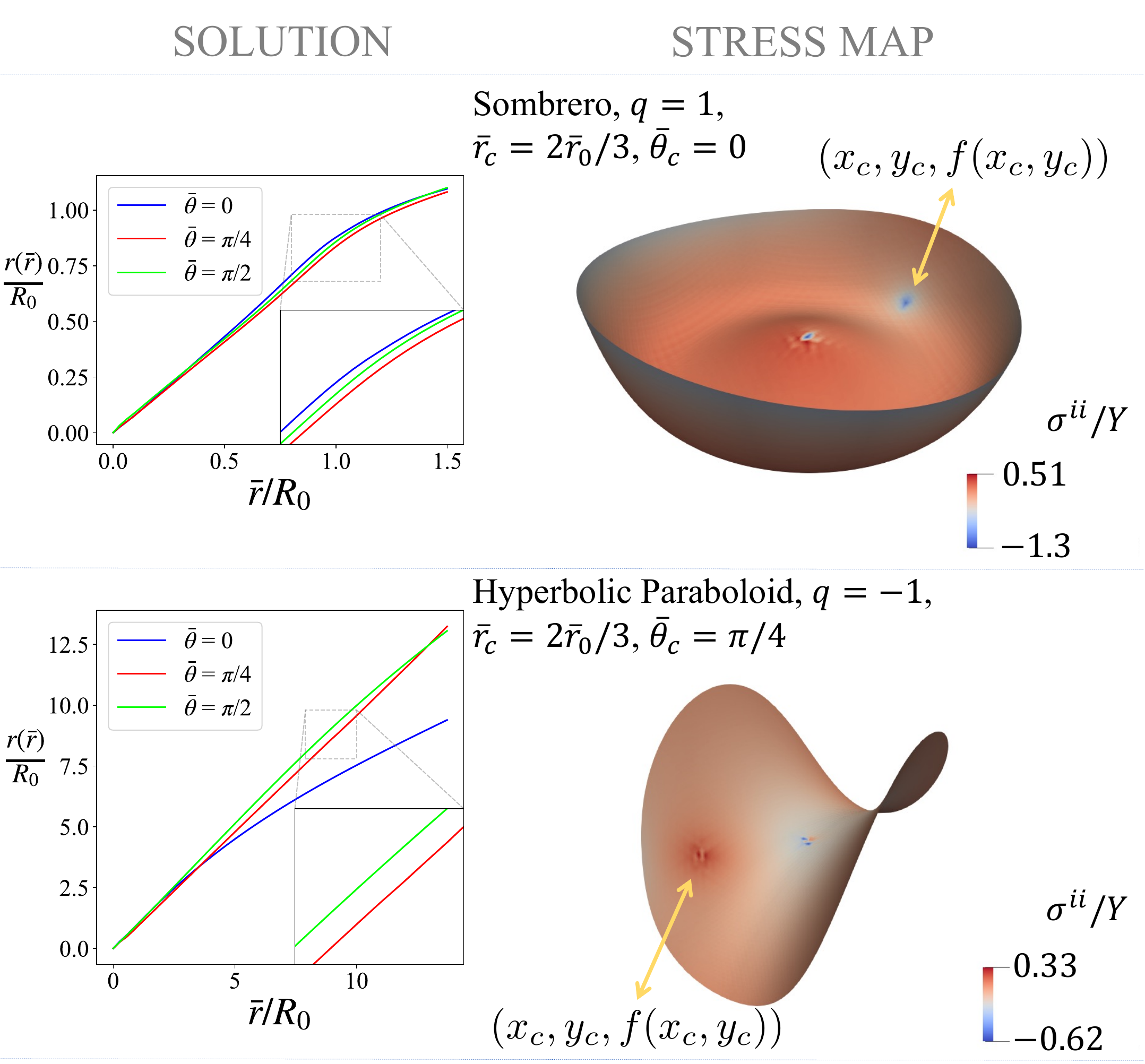}
    \caption{The first column shows the solutions $r(\bar{r})$ for a sombrero with parameters $(\bar{r}_0, \beta, R_0)=(15, 2, 10)$ and a hyperbolic paraboloid with parameters $(\bar{r}_0, \beta, R_0)=(112, 1/2, 8)$ at different angles, see the main text and Eq.~\eqref{Eq:results:rays}. The insets zoom in on the region near the disclination, highlighting its local effect. The right column shows the stress density, where stress is negative for a positive disclination and positive for a negative disclination. The figure is created using Paraview \cite{paraview}.}
    \label{main:fig:stress_map}
\end{figure}

For the structures in Fig.~\ref{main:fig:geometry}, the mapping 
in polar coordinates is as follows,
\begin{eqnarray}\label{Eq:results:rays}
    r = r(\bar{r},\bar{\theta}),\ \theta = \theta(\bar{r},\bar{\theta})\ .
\end{eqnarray}
The solutions with rotational symmetry {\it i.e.},~$r=r(\bar{r}),\theta=\bar{\theta}$ 
{are shown for} the case of a spheroid without a disclination, see Fig.~\ref{appendix:fig:spheroid_stress}.  On the first column of Fig.~\ref{main:fig:stress_map}, we plot the solution $r(\bar{r})$ for different values of $\bar{\theta}$. {As illustrated by the case of sombrero in the figure, even if the actual space is a surface of revolution, the symmetry is broken by the off-centered disclination {\it i.e.}} the solution depends on $\theta$. The insets in the figure emphasize the region surrounding the disclinations. {The asymmetric or $\theta$-dependent effect of the solution is more pronounced when the actual surface exhibits strong anisotropy, such as in the case of the hyperbolic paraboloid with $\beta=1/2$, where the curvature is sharper along the $x$-direction and flatter along the $y$-direction.} 
The solution $\mathbf{x}(\bar{\mathbf{x}})$ indeed reflects the symmetry of the system, as demonstrated in the top rows of Figs.~S2-S6, where colors indicate the magnitude of the vector field $\bm{x}(\bm{\bar{x}})$, {\it i.e}., $|\bm{x}|$. Each $\bm{\bar{x}}$ in the reference space corresponds to a position $\bm{x}$ in the actual space, depicted in the second rows of the corresponding figures.

We can also obtain the stress map distribution as shown in the second column of Fig.~\ref{main:fig:stress_map}. The positive disclination in the sombrero induces a local negative stress, while the negative disclination induces a local positive stress. This trend holds similarly for the hyperbolic paraboloid. The existence of the disclination completely distorts the stress map compared to the disclination-free case (Fig.~\ref{appendix:fig:spheroid_stress}). Figures~S7~and~S8 show the stress map in the actual space corresponding to the solutions presented in Figs.~S2-S6.


\begin{figure}
    \centering
    \includegraphics[width=\linewidth]{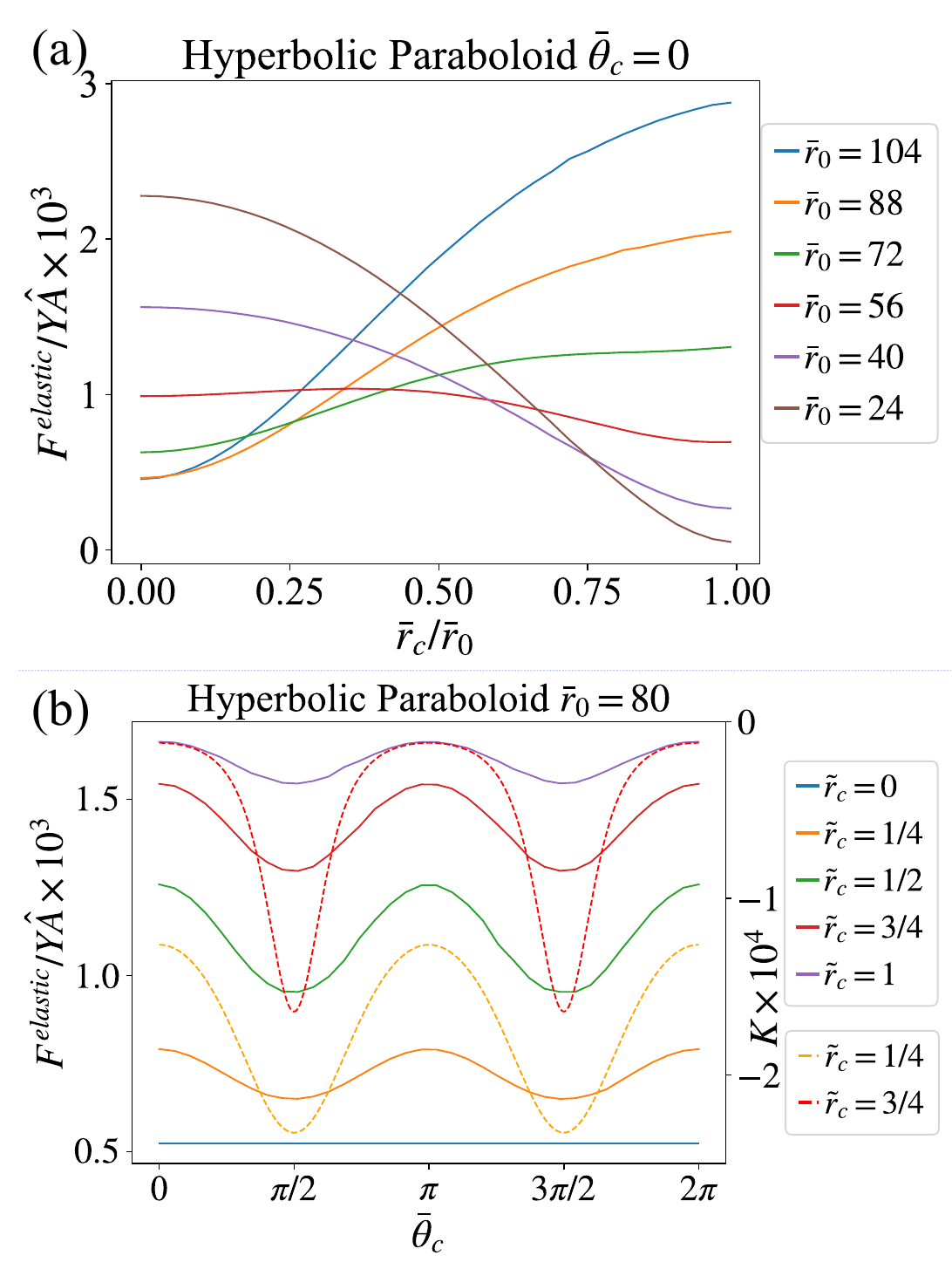}
    \caption{(a) Energy as a function of distance between the disclination and the reference domain center for a hyperbolic paraboloid with $\beta=1/2$, $R_0=8$ and $\bar{\theta}_c=0$ for different circular reference domain with radii $\bar{r}_0$, see the panel on the right. (b) The solid line shows the energy as a function of $\bar{\theta}_c$ for a given radius \add{$\tilde{r}_c \equiv \bar{r}_c/r_0$}. The radius of the circular reference domain is $\bar{r}_0 = 80$. The dashed lines represent the Gaussian curvature.}
    \label{main:fig:energy_as_dis_moves}
\end{figure}

To obtain the elastic energy of the system, we substitute the solutions of Eq.~\eqref{Eq:intro:GM:covariant} presented in Figs.~\ref{main:fig:stress_map} and S2-S6 into the expression for ${\cal F}^{elastic}$ and integrate over the reference domain using the metric $g(\bar{\bm{x}})$. Figure~\ref{main:fig:energy_as_dis_moves}a shows the elastic energy of a hyperbolic paraboloid for various sizes of circular domains in the reference space, plotted against the disclination position as it moves from the center $\bar{r}_c=0$ to the boundary at $\bar{r}_c=\bar{r}_0$ along a fixed angle $\bar{\theta}_c=0$ 
(see also Fig.~S9). 
The results show that for the small values of $\bar{r}_0/R_0$, the minimum of the elastic energy is achieved when the disclination is placed at the boundary, while for increasing $\bar{r}_0/R_0$ a disclination at the center becomes energetically favored, with an intermediate situation where the free energy is relatively insensitive to the position of the disclination. A sharper curvature, corresponding to the change from $\beta=1/2$ to $\beta=1$, facilitates this transition. The case of $\beta=1$ is illustrated in Fig.~\ref{appendix:fig:hyp_energy_dis_go_radial_beta_1}. For cases involving spheroids and sombreros see Fig.~S9b, c.  It is important to note that to validate the accuracy of our numerical method, we compared the solutions obtained using finite element methods with those from Ref. \cite{YinanMe2022} under conditions of rotational symmetry. Figure~S10 demonstrates a perfect match between them.

The hyperbolic paraboloid is not a surface of revolution, so the optimal disclination position depends on the angle $\bar{\theta}_c$ as well. This is illustrated in Fig.~\ref{main:fig:energy_as_dis_moves}b for $\beta=1/2$, where a disclination is positioned at different fixed radii $\bar{r}_c$, see also Fig.~S11 for different reference domain radius $\bar{r}_0$. Due to the $C_2$ symmetry of the surface, when the disclination moves in a circle around the axis of symmetry, the energy oscillates with a period of $\pi$. For $\beta \rightarrow 1$, the Gaussian curvature becomes rotationally invariant according to Eq.~(S4), and consequently, the energy becomes constant (Fig.~S12).




\begin{figure}
    \centering
    \includegraphics[width=.7\linewidth]{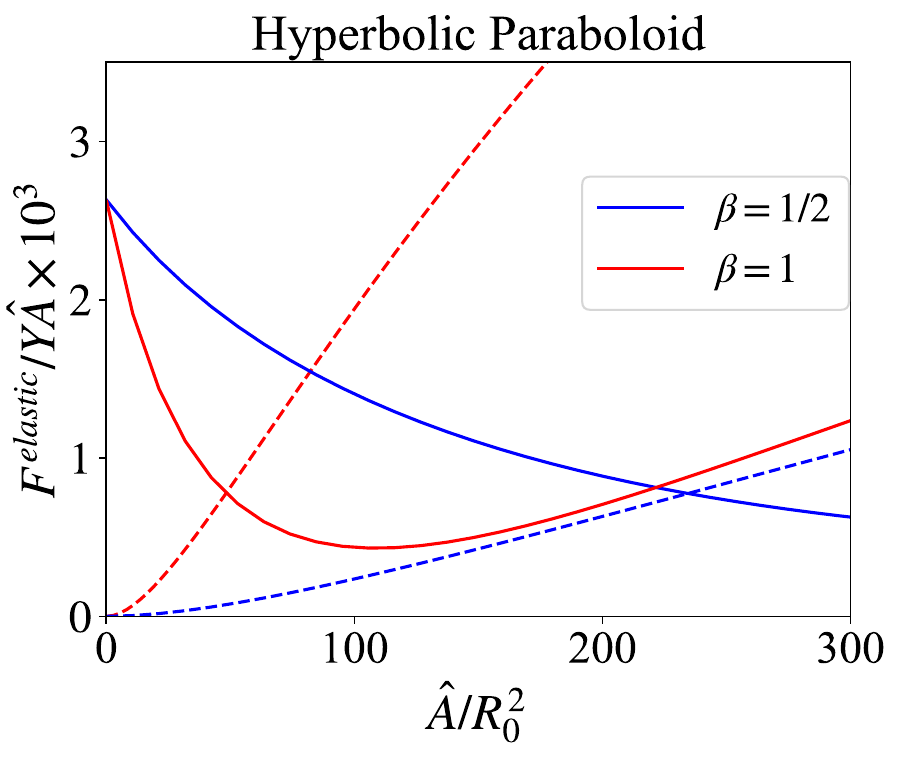}
    \caption{Energy vs.~the domain area in reference space for a square domain on a hyperbolic paraboloid with $R_0=8$ for $\beta=1/2,1$. The dashed lines represent the energy without a disclination, and the solid ones show energy with a centered 7-fold disclination. The intersection of the dashed and solid lines indicates the transition area where the formation of a discliantion at the center becomes energetically favorable.  The  intersection area values are listed in TABLE~\ref{main:table:hyp_onset_list}.}
    \label{main:fig:hyp_onset_square}
\end{figure}
\setcounter{figure}{0}
\renewcommand{\figurename}{TABLE}
\renewcommand{\thefigure}{\arabic{figure}}
\begin{figure}
    \centering
    \includegraphics[width=\linewidth]{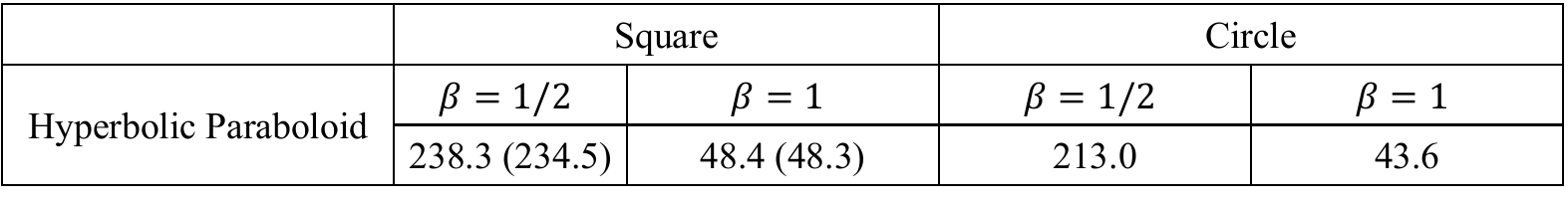}
    \caption{Transition normalized areas $A/R_0^2$ (see Fig.~\ref{main:fig:hyp_onset_square}) beyond which a disclination becomes energetically favorable for both square and circular domains on a hyperbolic paraboloid.  The values in brackets show the transition point when the square shown in Fig.~\ref{main:fig:geometry}b is rotated by $\pi/4$. The table highlights the difference between the transition areas of circular and square domains.}
    \label{main:table:hyp_onset_list}
\end{figure}

We also examined {the role of the boundary by considering a square domain in reference space}.  Figure~\ref{main:fig:hyp_onset_square} illustrates the elastic energy as a function of the area in reference {space} for a hyperbolic paraboloid with $\beta=1/2$ and $1$. Solid lines indicate configurations with a centered disclination, while dashed lines depict cases without, using a square reference domain. The solid lines in the figure show that the energy with a disclination located at the center decreases as the area increases, but only for very small areas due to the disclination-curvature interaction. As the domain area becomes sufficiently large, the energy then increases. In contrast, the elastic energy without a disclination increases monotonically. The intersection of the dashed and solid lines indicates the area at which a disclination becomes favorable. For comparison, similar curves are plotted for circular domains in Fig.~S14. The areas at these intersections are listed in Table.~\ref{main:table:hyp_onset_list} for a hyperbolic paraboloid surface with square and circle domains.  Clearly, the deviation of the reference domain from a circle to a square delays the onset of the first disclination, indicating the significant role of symmetry in crystal growth.  Figure S13 shows the elastic energy of a spheroid and a sombrero with a square domain. 


In summary, in this paper, we have conducted a comprehensive investigation of curvature-disclination interactions within the framework of full non-linear theory without any approximations. While {formulations} of non-linear elasticity theory in terms of geometric invariants has been previously developed\cite{Efrati2009, MosheSharonKupferman2015}, {exact solutions for arbitrary geometries were not previously available}~\cite{LiTravesset2019,Li2019,YinanMe2022}. In this paper, we overcame the challenge of lifting rotational symmetry restrictions and calculated the free energy of curved surfaces without any constraint on the disclination's location.

{Transitioning from one dimension to two dimensions is a significant step. This paper represents the first thorough examination of the interaction between a topological defect and a general manifold in the continuum limit.} 
{Previous simulations have shown that boundary fluctuations and line tension can shift the location of the first pentameric defect \cite{LiTravesset2018, Sanaz2020}, which is critical because the curvature induced by this initial disclination significantly affects the final structure of the shell, determining whether it forms a cylinder, sphere, or cone \cite{PRLZandi2009}. These effects were previously difficult to study due to the lack of a comprehensive theory that accounts for all relevant factors. Our method directly addresses these challenges and provides a powerful tool for further investigations.} To fully understand how 12 pentameric defects emerge in structures like HIV and other crystals during growth, it is essential 
{to determine the reference metric for a multi-disclination configuration.}. This is a formidable task, which we leave for future work.


\bigskip
The work of AT is funded by NSF, DMR-CMMT-2402548. RZ, YL and SL acknowledge support from NSF DMR-2131963 and the University of California Multicampus Research Programs and Initiatives (grant No. M21PR3267).
\newpage
\setcounter{figure}{0}
\setcounter{equation}{0}
\setcounter{table}{0}
\setcounter{section}{0}
\renewcommand{\thefigure}{A\arabic{figure}}
\renewcommand{\theequation}{A\arabic{equation}}
\renewcommand{\thetable}{A\arabic{table}}
\renewcommand{\figurename}{FIG.}
\newpage
\textit{Appendix:} In this appendix, we present {the details of the simulations performed in Ref.~\cite{nature2016}, the connection between the discrete and the continuous models,} the actual metric, the stress distributions map, and the solutions $r=r(\bar{r})$ along different directions on a spheroid without disclinations. We also show the energy as a function of the disclination's radial distance from the center of symmetry on a hyperbolic paraboloid with $\beta=1$.

{The conical and cylindrical structures presented in Fig.~1 are simulated by the addition of equilateral triangular subunits (trimers) to a growing shell. At each step of growth, a trimer is added at the location that minimizes the energy. The energy of the system can be expressed as follows:
\begin{align}
    F_d = E_s + E_b &= \sum_{i}\frac{1}{2} k_s (b_i - b_0)^2 \nonumber\\
    &+ \sum_{i,j} k_b [1 - \cos (\theta_{ij} - \theta_0)]\ .\label{Eq:discrete:energy}
\end{align}
with $k_s$ the stretching rigidity and $b_0$ the equilibrium bond length. The stretching energy $E_s$ sums over all bonds $i$. The bending energy $E_b$ sums over all neighboring trimers $i,j$ with $k_b$ the bending rigidity and $\theta_0$ the preferred angle between the two neighbors. The spontaneous curvature $H_0$ is related to the preferred angle by
\begin{align}\label{Eq:spontaneous:curvature:preferred:angle}
    H_0^2 b_0^2 = \frac{12 \sin^2 (\theta_0/2)}{1 + 3 \sin^2 (\theta_0/2)}\ .
\end{align}
A critical step during shell growth is the formation of pentamers or hexamers. Depending on which configuration has the lower energy, five triangles may join to form a pentamer, or a sixth triangle may be added to form a hexamer (see Ref.\cite{LiTravesset2018,panahandeh2022virus} for more details). If pentamers are not favored at any point during growth, the shell forms a cylindrical structure; otherwise, a spherical, conical, or completely irregular structure \cite{li2022biophysical} may develop.

The elastic energy density 
\begin{align}
    \mathcal{F}^{elastic}_{linear} = \frac{1}{2} (2\mu u_{\alpha \beta}^2 + \lambda u_{\alpha \alpha}^2)\label{eq:linear:elastic:energy:density}
\end{align}
is obtained by taking the continuum limit of the stretching energy in Eq.~\eqref{Eq:discrete:energy} with the Lam\'e coefficients $\mu = \lambda = \sqrt{3} \epsilon/4$, $\epsilon = k_s/b_0$ \cite{Seung1988}. Equation~\eqref{eq:linear:elastic:energy:density} represents the linear order expansion of the covariant elastic energy density, which is utilized in this paper in Eqs.~\eqref{Eq:intro:free_dens} and (S1) \cite{YinanMe2022}. The continuum limit of the bending energy in Eq.~\eqref{Eq:discrete:energy} corresponds to $\mathcal{F}^{bending}$ presented in the paragraph below Eq.~\eqref{Eq:intro:free_dens}. In this paper, we minimize $\mathcal{F}^{elastic}$ for a given fixed geometry.

}

For the surfaces presented in Fig.~\ref{main:fig:geometry}, we employ the Monge representation\cite{DESERNO201511} to describe the location of the points in actual space, given by $\bm{x}=(x,y,f(x,y))$. The position of the disclination is denoted as $(x_c,y_c,f(x_c,y_c))$. The actual metric, then, reads
\begin{equation}\label{Eq:intro:actual_metric}
    g_{\mu\nu}=\begin{bmatrix}
    1 + (\partial_xf)^2 & (\partial_xf) (\partial_yf)\\
    (\partial_xf) (\partial_yf) & 1 + (\partial_yf)^2
    \end{bmatrix}\ .
\end{equation}

\begin{figure}[H]
    \centering
    \includegraphics[width=.45\textwidth]{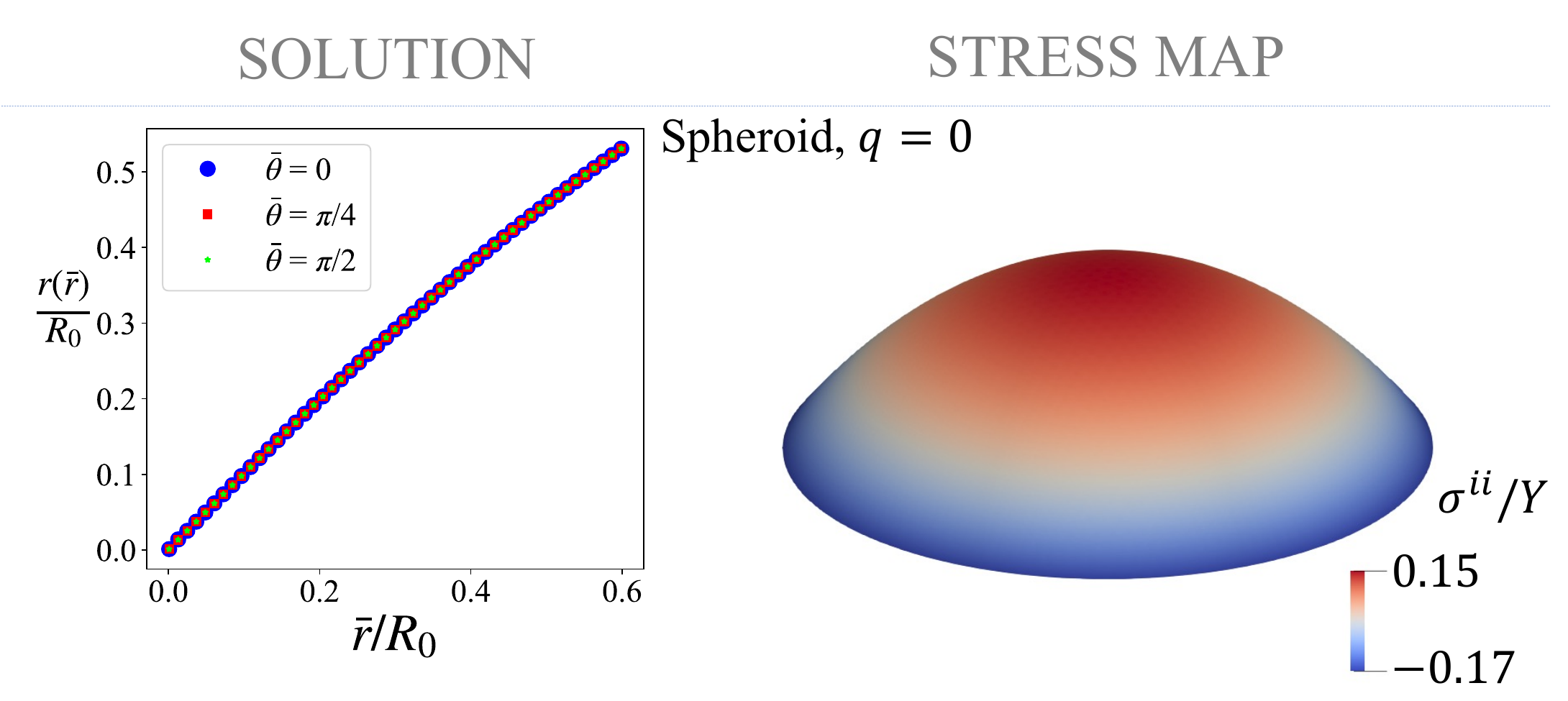}
    \caption{The left column shows the solution $\bm{x}(\bar{\bm{x}})$ for a spheroid ($(\bar{r}_0, \beta, R_0)=(6, 2, 10)$) without any disclinations for various angles $\bar{\theta}$, see Eq.~\eqref{Eq:results:rays} and the related discussion. All the plots overlap due to the system's symmetry. The right column displays the stress map distribution, which is positive at the center and negative at the edge.}
    \label{appendix:fig:spheroid_stress}
\end{figure}

\begin{figure}[H]
    \centering
    \includegraphics[width=.45\textwidth]{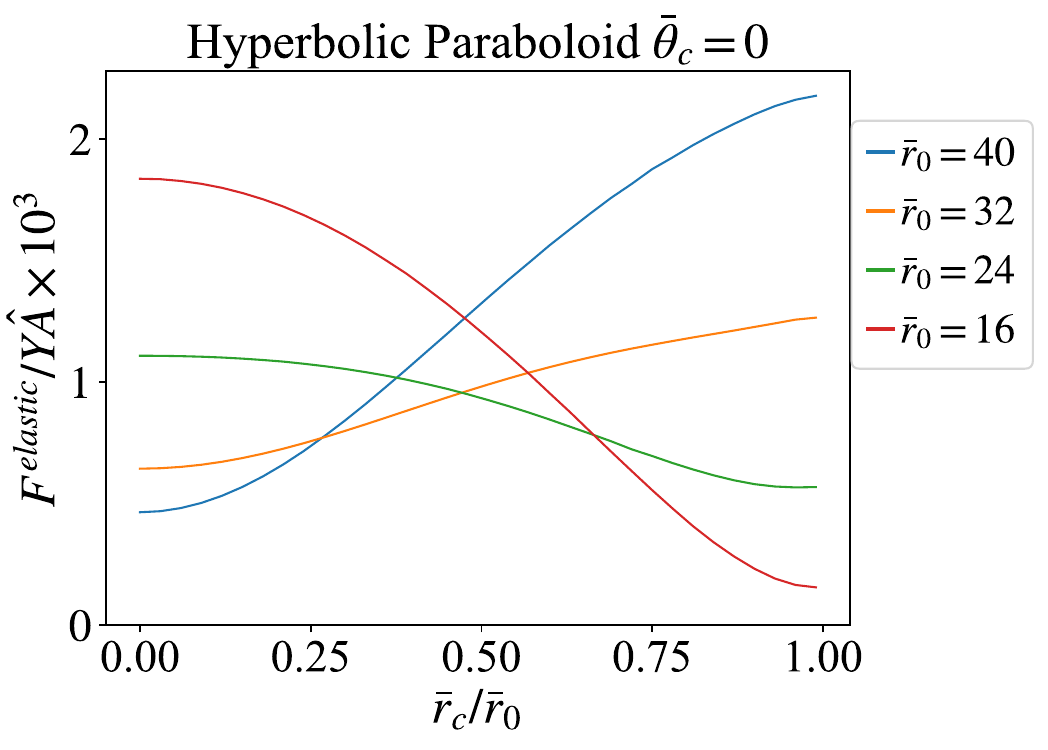}
    \caption{Energy as a function of the distance between the disclination and the reference domain center for a hyperbolic paraboloid with $\beta=1$, $R_0=8$ and $\bar{\theta}_c=0$ for different circular reference domain with radii $\bar{r}_0$, see the panel on the right. For larger areas, the formation of a disclination at the center becomes energetically more favorable. Conversely, as the area decreases, the system's energy is minimized when the disclination moves to the edge.
    }
    \label{appendix:fig:hyp_energy_dis_go_radial_beta_1}
\end{figure}


\bibliography{references.bib, references_liu}

\clearpage
\onecolumngrid
\setcounter{figure}{0}
\setcounter{equation}{0}
\setcounter{table}{0}
\setcounter{section}{0}
\renewcommand{\thefigure}{S\arabic{figure}}
\renewcommand{\theequation}{S\arabic{equation}}
\renewcommand{\thetable}{S\arabic{table}}
\renewcommand{\figurename}{FIG.}
\setcounter{page}{1}

\centerline{\bf Supplementary Information}

\begin{center} 
	{\bf General Solution for Elastic Networks on Arbitrary Curved Surfaces in the Absence of Rotational Symmetry}\\
	\bigskip
	\normalsize
	Yankang Liu,$^1$ Siyu Li,$^1$, Roya Zandi,$^1$ and Alex Travesset,$^2$ \\
	\bigskip
	{$^1$\it Department of Physics and Astronomy,\\
		University of California Riverside, Riverside, California 92521, United States}\\
	{$^2$\it Department of Physics and Astronomy and Ames Lab,\\
		Iowa State University, Ames, Iowa 50011, United States}\\
\end{center}
\bigskip
\section{\label{app:A}Explicit Formulas for the Different Quantities}
In this appendix, we provide explicit expressions for the formulas in the main text.

The explicit form of ${\cal F}^{elastic}$ is
\begin{equation}\label{Eq:app:elastic energy density}
    {\cal F}^{elastic}=\frac{1}{2} A^{\alpha\beta\gamma\delta} u_{\alpha \beta} u_{\gamma \delta},
\end{equation}
where $u_{\alpha\beta}$ (see main text) is the strain tensor and 
\begin{equation}\label{Eq:app:EF:Elastic_constant}
    A^{\alpha\beta\gamma\delta}=\frac{Y}{1-\nu^2_p}\left[\nu_p g^{\alpha\beta}g^{\gamma\delta}+(1-\nu_p)g^{\alpha\gamma}g^{\beta\delta} \right] \ ,
\end{equation}
with $Y$ the Young Modulus, $\nu_p$ the Poisson ratio and $g^{\alpha\beta}$ the inverse of the actual metric such that $g^{\mu\nu}g_{\nu\gamma}=g_{\nu\gamma}g^{\mu\nu}=\delta^{\mu}_{\gamma}$ where $\delta^{\mu}_{\gamma}$ is the Kronecker delta function. The stress tensor is
\begin{equation}\label{Eq:app:EF:Stress}
    \sigma^{\alpha\beta}=\frac{1}{\sqrt{g}}\frac{\delta F}{\delta u_{\alpha \beta}}=A^{\alpha\beta\gamma\delta}  u_{\gamma \delta} \ .
\end{equation}

The Gaussian curvature is
\begin{equation}\label{Eq:app:EF:gaussian}
    K = \frac{\mbox{det}(\partial_i\partial_j f)}{(1+(\nabla f)^2)^2}=\frac{f^{\prime}(r)f^{\prime\prime}(r)}{r(1+f^{\prime}(r)^2)^2} \ ,
\end{equation}
and the mean curvature (with the convention that $R_i=R>0,i=1,2$ for the sphere) is
\begin{eqnarray}\label{Eq:app:EF:Mean}
    2H &=& -\nabla \cdot \left( \frac{\nabla f}{(1+(\nabla f)^2)^{1/2}} \right)\\ \nonumber
    &=&-\left( \frac{f^{\prime\prime}(r)}{(1+f^{\prime}(r)^2)^{3/2}}+\frac{f^{\prime}(r)}{r(1+f^{\prime}(r)^2)^{1/2}} \right)
\end{eqnarray}
see Ref.~\cite{Zandi2020} for the details. The two curvatures can be obtained from the equation
\begin{eqnarray}
    K&=&\frac{1}{R_1 R_2} \quad \nonumber\\
    2H&=&\frac{1}{R_1}+\frac{1}{R_2} \ ,
\end{eqnarray}
such that $\frac{1}{R_1}=H+\sqrt{H^2-K}$ and $\frac{1}{R_2}=H-\sqrt{H^2-K}$ with $H$ and $K$ given in Eqs.~\eqref{Eq:app:EF:gaussian} and ~\eqref{Eq:app:EF:Mean}.

Under a change of variables ${\bm y}({\bm x})$, the metric becomes
\begin{eqnarray}
    g'_{\mu \nu}({\bm y}) &=&\frac{\partial x^{\alpha}}{\partial y^{\mu}} \frac{\partial x^{\beta}}{\partial y^{\nu}} g_{\alpha \beta}({\bm x})\ .
\end{eqnarray}
The determinant of the metric becomes
\begin{eqnarray}
    \sqrt{g'}({\bm y}) &=&\left| \frac{\partial x^{\alpha}}{\partial y^{\mu}} \right| \sqrt{g}({\bm x})\ .
\end{eqnarray}
And the surface element becomes
\begin{eqnarray}
    d^2{\bm y} &=& \left| \frac{\partial y^{\alpha}}{\partial x^{\mu}} \right| d^2{\bm x}\ .
\end{eqnarray}


\section{\label{app:C}Boundary Condition}

The boundary conditions can be obtained through the variations of $F^{area} = F^{elastic} + F^{bending}$ in Eq.~\eqref{Eq:intro:free_dens} and the reparameterizations of the actual metric,
\begin{equation}\label{Eq:variation:reparam_ractual}
    \delta F^{area} = -\int d^2{\bm x}\partial_{\rho}\left(\sqrt{g}\sigma^{\rho \mu}\bar{\xi}_{\mu}\right) =\oint \sqrt{g} dx^{\mu}\varepsilon_{\mu\nu} \sigma^{\nu \rho} \bar{\xi}_{\rho} \ .
\end{equation}

If there is a line tension $\tau$, the contribution to the free energy is then
\begin{equation}\label{Eq:app:line energy}
    F^{line} =\tau \oint ds = \tau \oint \sqrt{\langle g \rangle} dl=\tau \oint dx^{\mu} g_{\mu\nu} t^{\nu} \ ,
\end{equation}
where $\langle g \rangle=g_{\mu\nu} \frac{d x^{\mu}}{dl} \frac{d x^{\nu}}{dl}$ and
\begin{equation}
    t^{\mu}=\frac{1}{\sqrt{\langle g \rangle}} \frac{dx^{\mu}}{dl},
\end{equation}
is the unit vector (in the actual metric) tangent to the boundary curve with $dx^{\mu}=\sqrt{\langle g \rangle} t^{\mu} dl$. The variations to the free energy Eq.~\eqref{Eq:app:line energy} as shown in Ref.~\cite{LiTravesset2019} then become,
\begin{eqnarray}
    \delta F^{line}&=&\frac{\tau}{2} \oint dx^{\mu} \delta g_{\mu \nu} t^{\nu} =- \tau  \oint dx^{\mu} \nabla_{\mu}\xi_{\nu} t^{\nu}\\\nonumber
    &=&\tau  \oint dx^{\mu} \xi_{\nu} \nabla_{\mu} t^{\nu} 
\end{eqnarray}
Since $\delta\left(F^{area}+F^{line}\right)=0$, the appropriate boundary condition is then,
\begin{equation}
    \tau g_{\rho \nu} \nabla_{\mu} t^{\rho} = -\sqrt{g} \epsilon_{\mu \rho} \sigma^{\rho \lambda} \bar{g}_{\lambda \nu} \ .
\end{equation}
Note the tangent vector is 
\begin{equation} \label{Eq:tangentvector}
    t^{\rho}\nabla_{\rho} t^{\mu} =-\frac{1}{r_A} n^{\mu} \ ,
\end{equation}
where $r_A=-1/(\sqrt{g}\epsilon_{\mu\nu}t^{\nu}t^{\rho}\nabla_{\rho}t^{\mu})$ is the curvature of the curve defining the boundary. The normal to the tangent vector can be written as
\begin{equation}
     n_{\mu} = \sqrt{g} \varepsilon_{\mu \rho} t^{\rho}. 
\end{equation}
The boundary condition, thus, becomes
\begin{equation}
    n_{\rho}\sigma^{\rho \lambda}\bar{g}_{\lambda \nu} = -\frac{\tau}{r_A} n_{\nu} \ .
\end{equation}

As an example, if the problem has rotational symmetry and the boundary is a circle $r=r_0$, then $\theta(s)=s/r_0$.  The tangent vectors then are 
\begin{equation}
    t^{r}=0 \quad t^{\theta}=\frac{1}{r_0},
\end{equation}
and the normal vectors
\begin{equation}
    n_r = \sqrt{1+f^{\prime}(r_0)^2} \quad n_{\theta}=0.
\end{equation}
Then Eq.~\eqref{Eq:tangentvector} becomes
\begin{eqnarray}
    t^{\rho}\nabla_{\rho} t^{r}&=&\Gamma^{r}_{\theta\theta}\frac{1}{r^2_0}=-\frac{1}{r_0(1+f^{\prime}(r_0)^2)} \\\nonumber
    &=& -\frac{1}{r_0\sqrt{1+f^{\prime}(r_0)^2}} n^{r}
\end{eqnarray}
with 
\begin{equation}
    \frac{r_A}{r_0}=\sqrt{1+f^{\prime}(r_0)^2}.
\end{equation}
Finally, the boundary condition becomes equal to,
\begin{equation}\label{Eq:app:boundary condition with rotational symmetry}
    \overline{g}_{rr}(r_0)\sigma^{rr}(r_0)=-\frac{1}{\sqrt{1+f^{\prime}(r_0)^2}}\frac{\tau}{r_0} \ .
\end{equation}
For a sphere $f^{\prime}(r)=-\frac{r}{\sqrt{R^2_0-r^2}}$ and thus
\begin{equation}
    r_A = \frac{r_0}{\sqrt{1-\left(\frac{r_0}{R_0}\right)^2}} \ .
\end{equation}

\section{Free energy Normalization}\label{app:normalization}

We will consider a dimensionless free energy normalized per particle, that is
\begin{equation}\label{Eq:SI:normalization:free}
    f\equiv \frac{F}{Y \bar{A}}=\frac{2F}{\sqrt{3} N Y a^2_L} \ ,
\end{equation}
hence the area in {\em reference space}, see Eq.~\eqref{Eq:intro:area_reference}. This area is given by
\begin{equation}
\hat{A}=\pi\left(1-\frac{q_i}{6}\right) \rho_0^2 \ .
\end{equation}
Given two systems with the same number of particles, the one with the smallest free energy per particle, Eq.~\eqref{Eq:SI:normalization:free} is the stable minimum.

\section{About units}\label{app:units}

The free energy, see Eq.~\eqref{Eq:intro:free_dens} is
\begin{equation}\label{Eq:units:tot}
     F = \int d^2{\bm x}\sqrt{g}\left[{\cal F}^{elastic} + {\cal F}^{bending}\right]+F^{abs}+F^{line} \ .
\end{equation}

Note that the stress tensor, Eq.~\eqref{Eq:app:EF:Stress} is given by
\begin{equation}
    \sigma^{\alpha \beta}= \frac{1}{\sqrt{g}}\frac{\delta F}{\delta u_{\alpha \beta}}=A^{\alpha\beta\gamma\delta}  u_{\gamma \delta} = Y \times \left(\mbox{Terms that Depend on } \nu_p\right)
\end{equation}
Note also, that the ratio of the Young modulus and the line tension defines a coefficient $l_A$ with units of length 
\begin{equation}
    \frac{\tau}{Y}\equiv l_A
\end{equation}
Therefore, through the boundary conditions the quantity
\begin{equation}
    \frac{\sigma^{\alpha\beta}}{Y}=h(\nu_p,l_A) \ ,
\end{equation}
does not directly depend on the Young modulus. 

Also,
\begin{equation}
    F^{abs}=-N \Delta F = - \frac{2\Delta F}{\sqrt{3}a^2_L} \frac{\sqrt{3}}{2} N a^2_L =-\Pi \hat{A} 
    \mbox{ with } \hat{A}=\int d^2{\bm r}\sqrt{\bar{g}({\bm x})} \ .
\end{equation}
Therefore $\Pi=\frac{2\Delta F}{\sqrt{3}a^2_L}$, and $\hat{A}$ is the area in reference space.

The line tension term is a function of the perimeter ($P$), given by
\begin{equation}
    \frac{F^{line}}{YL^2}=\frac{\tau}{Y L^2} \oint_{\partial D} ds \equiv \frac{\tau P}{Y L^2} \ .
\end{equation}

Finally, the free energy Eq.~\eqref{Eq:units:tot} is
\begin{equation}
    \frac{F}{L^2Y}=f(\nu_p,\frac{l_A}{L})+\frac{\kappa}{YL^2}\int d^2{\bm x}\sqrt{g}\left[ \left(\frac{1}{R_1}-H_0\right)^2+ \left(\frac{1}{R_2}-H_0\right)^2\right]+\frac{\Pi}{Y}\frac{\hat{A}}{L^2} + \frac{\tau P}{Y L^2} \ ,
\end{equation}
where $L$ is a characteristic dimension of the system. 
In general we will choose $L^2=\hat{A}$, so
\begin{equation}
     \frac{F}{Y\hat{A}}=f(\nu_p,\frac{l_A}{\sqrt{\hat{A}}})+\frac{\kappa}{Y\hat{A}}\int d^2{\bm x}\sqrt{g}\left[ \left(\frac{1}{R_1}-H_0\right)^2+ \left(\frac{1}{R_2}-H_0\right)^2\right]+\frac{\Pi}{Y} + \frac{\tau P}{Y \hat{A}} \ ,
\end{equation}
which defines the effective linear tension $\hat{\tau}=\frac{\tau a_L}{Y \hat{A}}$ and dimensionless area $\hat{A}/a_L^2$, so that all lengths are expressed in terms of the lattice constant $a_L$.

\section{\label{SI:linear}Connection with Linear Elasticity Theory}

Here we show that the covariant formalism in the main text reduce to the known formulas from elasticity theory when the displacements are small. Within elasticity theory, the reference metric (without disclinations)
\begin{equation}\label{Eq:SI:LE:reference}
    {\bar g}_{\alpha\beta} = \delta_{\alpha \beta}
\end{equation}
The surface is described in the Monge gauge,
\begin{equation}
    z = h(x,y) \ .
\end{equation}
The mapping ${\bm x}={\cal U}({\bar x})$ is given by
\begin{equation}\label{Eq:SI:LE:mapping}
    {\bm x}=\bar{{\bm x}}+{\bm u}(\bar{{\bm x}}) \ ,
\end{equation}
where ${\bm u}$ is the displacement. Then, the actual metric becomes
\begin{eqnarray}\label{Eq:SI:LE:actual}
    g_{\alpha\beta}(\bar{\bm x}) &=& \bar{\partial}_{\alpha} {\vec r}({\bar {\bm x}}) \bar{\partial}_{\beta} {\vec r}({\bar {\bm x}})=\delta_{\alpha\beta}+\bar{\partial}_{\alpha} u_{\beta}+\bar{\partial}_{\beta} u_{\alpha}+ \bar{\partial}_{\alpha} u_{\gamma} \bar{\partial}_{\beta} u_{\gamma}+{\bar \partial}_{\rho} h {\bar \partial}_{\gamma} h\left(\delta_{\alpha\rho}\delta_{\gamma\beta}+\delta_{\alpha\rho}\bar{\partial}_{\beta}u_{\gamma}+\bar{\partial}_{\alpha}u_{\rho}\delta_{\beta\lambda}+ \bar{\partial}_{\alpha}u_{\rho}\bar{\partial}_{\beta}u_{\gamma}\right)
    \nonumber\\
     &\approx& \delta_{\alpha\beta}+\bar{\partial}_{\alpha} u_{\beta}+\bar{\partial}_{\beta} u_{\alpha}+{\bar \partial}_{\alpha} h {\bar \partial}_{\beta} h
\end{eqnarray}
If only linear terms in ${\bm u}$ and the leading term in $h$ are kept the
strain tensor $u_{\alpha\beta}$ becomes
\begin{equation}\label{Eq:SI:LE:strain}
    u_{\alpha \beta}=\frac{1}{2}\left(\bar{\partial}_{\alpha} u_{\beta}+\bar{\partial}_{\beta} u_{\alpha}+{\bar \partial}_{\alpha} h {\bar \partial}_{\beta} h \right) \ .
\end{equation}
The actual metric is
\begin{equation}
    g_{\alpha\beta}({\bar{\bm x}})=\bar{g}_{\alpha\beta}({\bar{\bm x}})+2 u_{\alpha \beta}({\bar{\bm x}})
\end{equation}
The leading elastic part of the free energy, consistent with the expansion Eq.~\eqref{Eq:SI:LE:strain} becomes
\begin{eqnarray}\label{Eq:SI:LE:free_energy}
{\cal F}^{elastic}&=&\frac{1}{2}\frac{Y}{1-\nu_p^2}\left(\nu_p(u_{\alpha\alpha}^2+(1-\nu_p)u_{\alpha\beta} u_{\alpha\beta} \right)
\nonumber \\
&=& \frac{1}{2} \left( 2\mu (u_{\alpha\beta})^2+\lambda (u_{\alpha\alpha})^2\right) 
\end{eqnarray}
expressed in terms of the Lame coefficients $\lambda,\mu$ instead of the Young modulus $Y=\frac{4\mu(\mu+\lambda)}{2\mu+\lambda}$ and Poisson ratio $\nu_p=\frac{\lambda}{2\mu+\lambda}$. For fixed geometry, that is for a given $f$, this is exactly the same free energy and strains as used in linear elasticity theory, see for example Ref.~\cite{Seung1988}. The Airy function is the solution to the equation, 
\begin{equation}\label{Eq:SI:LE:Airy}
    \frac{1}{Y_0}{\bar{\Delta}}^2 \chi(\bar{{\bm x}})=s({\bar{\bm x}})-K({\bar{\bm x}}) \ ,
\end{equation}
see Ref.~\cite{Seung1988} for a full derivation. The laplacian ${\bar{\Delta}}$ refers to a flat metric. Adding an arbitrary disclination density is done by introducing singularities in the reference metric, as discussed for a central disclination in the main text. It is
\begin{equation}\label{Eq:SI:LE:disc_density}
s({\bar{\bm x}})=\frac{\pi}{3}\sum_{i=1}^N q_i \delta({\bar{\bm x}}-{\bar{\bm x}}_i)
\end{equation}
where ${\bar x}_i$ are the positions of the $N$ disclinations. The Gaussian curvature is obtained by expanding Eq.~\eqref{Eq:app:EF:gaussian} to leading order, consistent with the expansion in the actual metric Eq.~\eqref{Eq:SI:LE:actual}. Therefore
\begin{equation}\label{Eq:SI:LE:Gaussian_curvature}
    K({\bar{\bm x}}) = -\frac{1}{2}\varepsilon_{\alpha \beta}\varepsilon_{\gamma \rho} \bar{\partial}_{\beta}  \bar{\partial}_{\rho} \left(\bar{\partial}_{\alpha} h \bar{\partial}_{\gamma} h\right) \ .
\end{equation}

We remark that Eq.~\eqref{Eq:SI:LE:Airy} is written in terms of a flat metric, and the only contribution from the curved surface is through the approximated Gaussian curvature Eq.~\eqref{Eq:SI:LE:Gaussian_curvature}. Eq.~\eqref{Eq:SI:LE:Airy} has the physical interpretation of the Gaussian curvature screening the disclination density

\section{\label{SI:plots}Additional plots}
\begin{figure}[H]
    \centering
    \includegraphics[width=.6\linewidth]{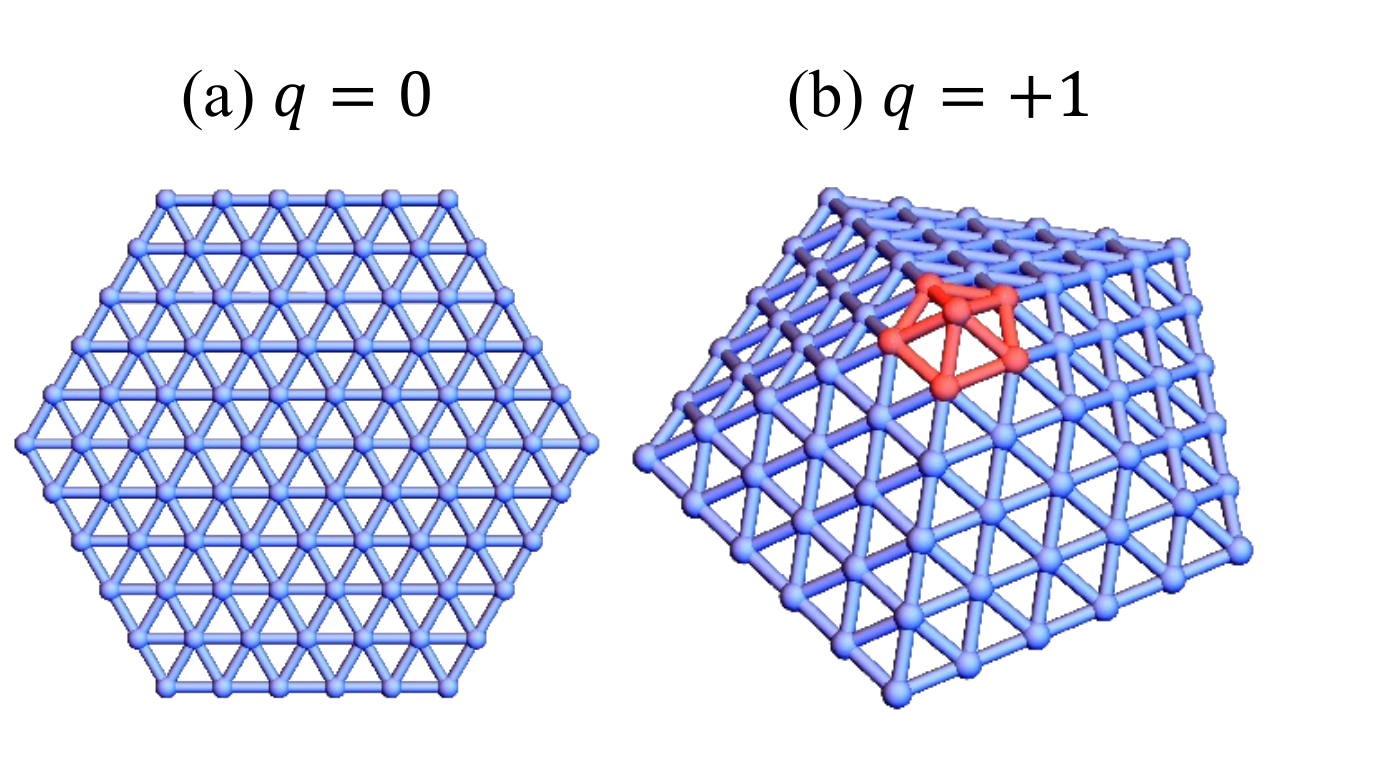}
    \caption{The lattice in the reference space consists of the equilateral triangles for $q=0, +1$. }
    \label{si:fig1}
\end{figure}

\begin{figure}
    \centering
    \includegraphics[width=\textwidth]{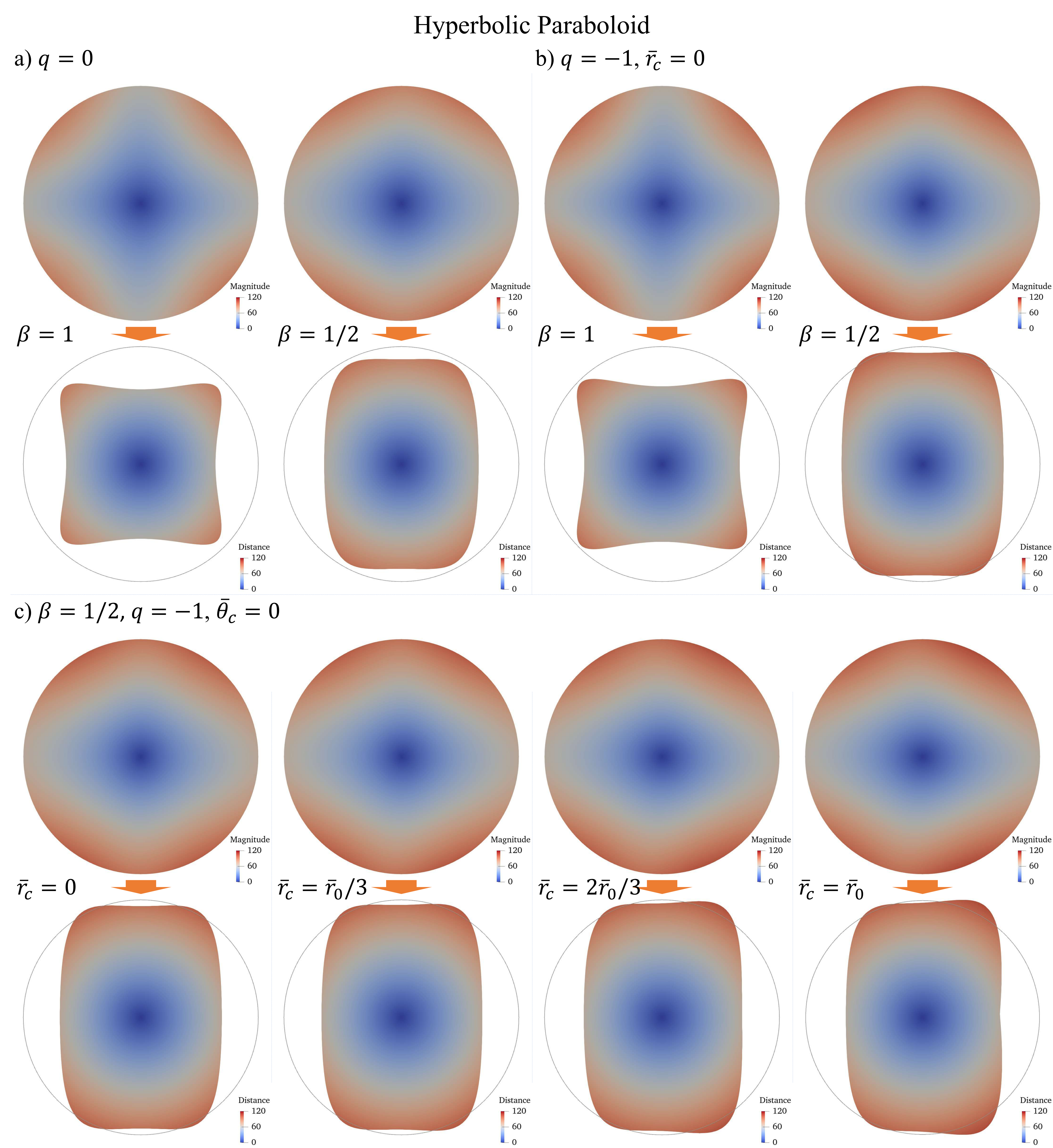}
    \caption{Solutions and corresponding actual domains for hyperbolic paraboloid. The upper rows present the solutions on the reference domain, where the color indicates the magnitude of the vector field $\bm{x}(\bm{\bar{x}})$, i.e., $|\bm{x}|$. With the solution, we can relocate each point $\bm{\bar{x}}$ in the reference space to its corresponding position $\bm{x}$ in the actual space, which constitutes the lower rows. Here, the color indicates the distance between that point to the $z$-axis. The circles in the lower rows circumscribe the reference domains for comparison. The position of the disclination in the reference space is specified using the polar coordinate. $(\bar{r}_0, R_0)=(96, 8)$.}
    \label{si:fig:3}
\end{figure}

\begin{figure}
    \centering
    \includegraphics[width=\textwidth]{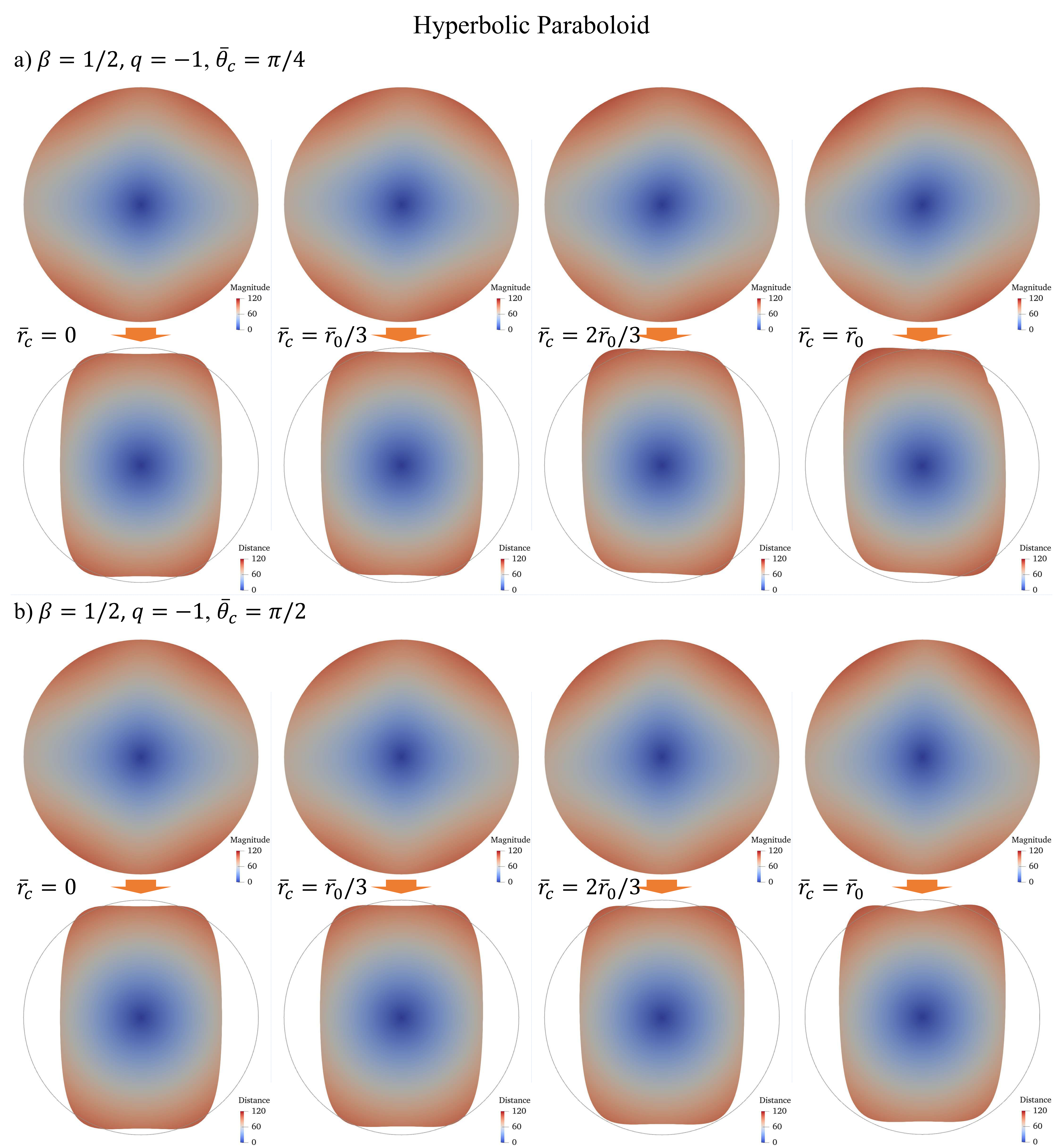}
    \caption{Solutions and corresponding actual domains for hyperbolic paraboloid. The upper rows present the solutions on the reference domain, where the color indicates the magnitude of the vector field $\bm{x}(\bm{\bar{x}})$, i.e., $|\bm{x}|$. With the solution, we can relocate each point $\bm{\bar{x}}$ in the reference space to its corresponding position $\bm{x}$ in the actual space, which constitutes the lower rows. Here, the color indicates the distance between that point to the $z$-axis. The circles in the lower rows circumscribe the reference domains for comparison. The position of the disclination in the reference space is specified using the polar coordinate. $(\bar{r}_0, R_0)=(96, 8)$.}
    \label{si:fig:4}
\end{figure}

\begin{figure}
    \centering
    \includegraphics[width=\textwidth]{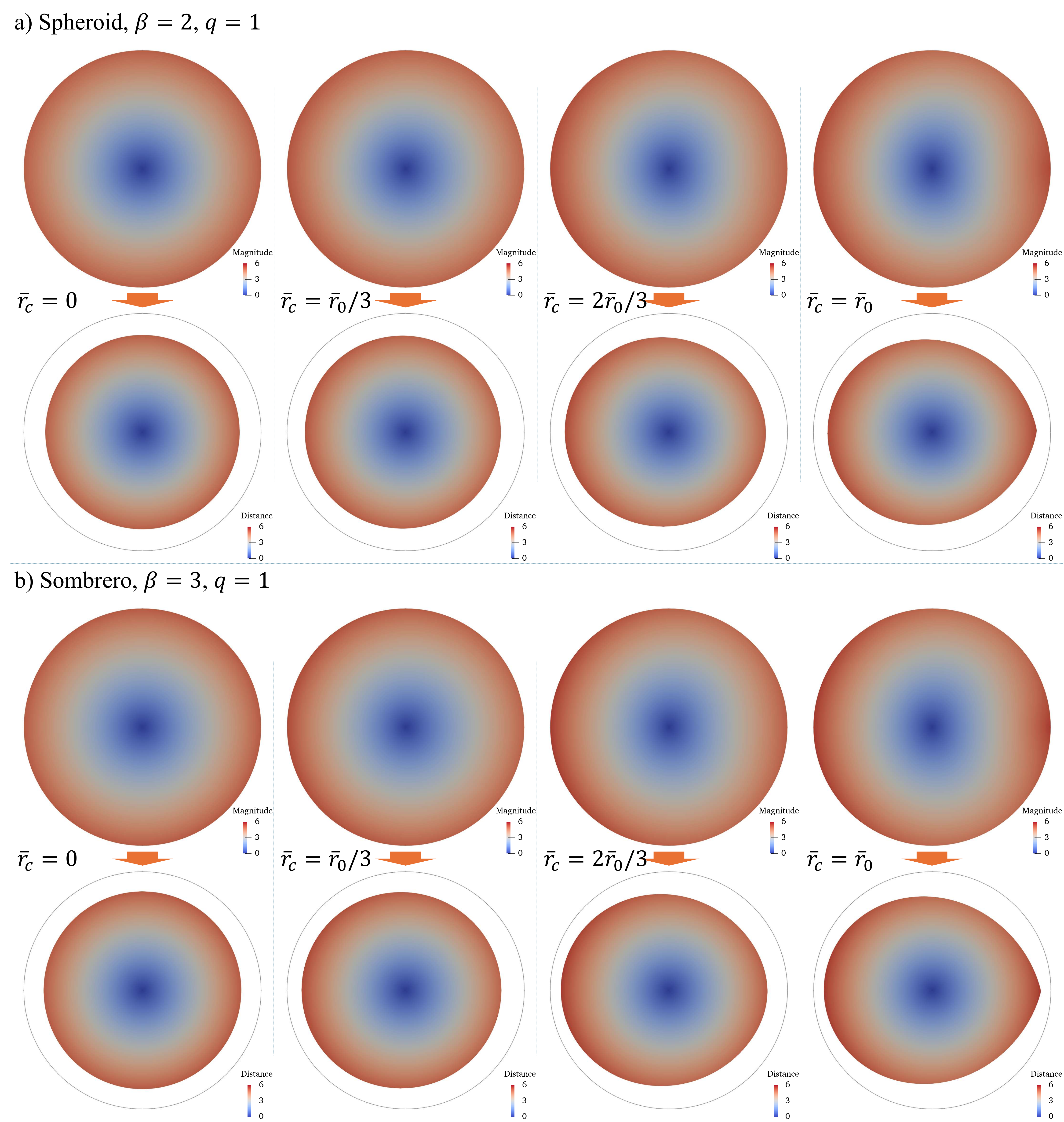}
    \caption{Solutions and corresponding actual domains for spheroids (a) and sombrero (b). The upper rows present the solutions on the reference domain, where the color indicates the magnitude of the vector field $\bm{x}(\bm{\bar{x}})$, i.e., $|\bm{x}|$. With the solution, we can relocate each point $\bm{\bar{x}}$ in the reference space to its corresponding position $\bm{x}$ in the actual space, which constitutes the lower rows. Here, the color indicates the distance between that point to the $z$-axis. The circles in the lower rows circumscribe the reference domains for comparison. The position of the disclination in the reference space is specified using the polar coordinate. $(\bar{r}_0, R_0)=(6, 10)$.}
    \label{si:fig:5}
\end{figure}

\begin{figure}
    \centering
    \includegraphics[width=\textwidth]{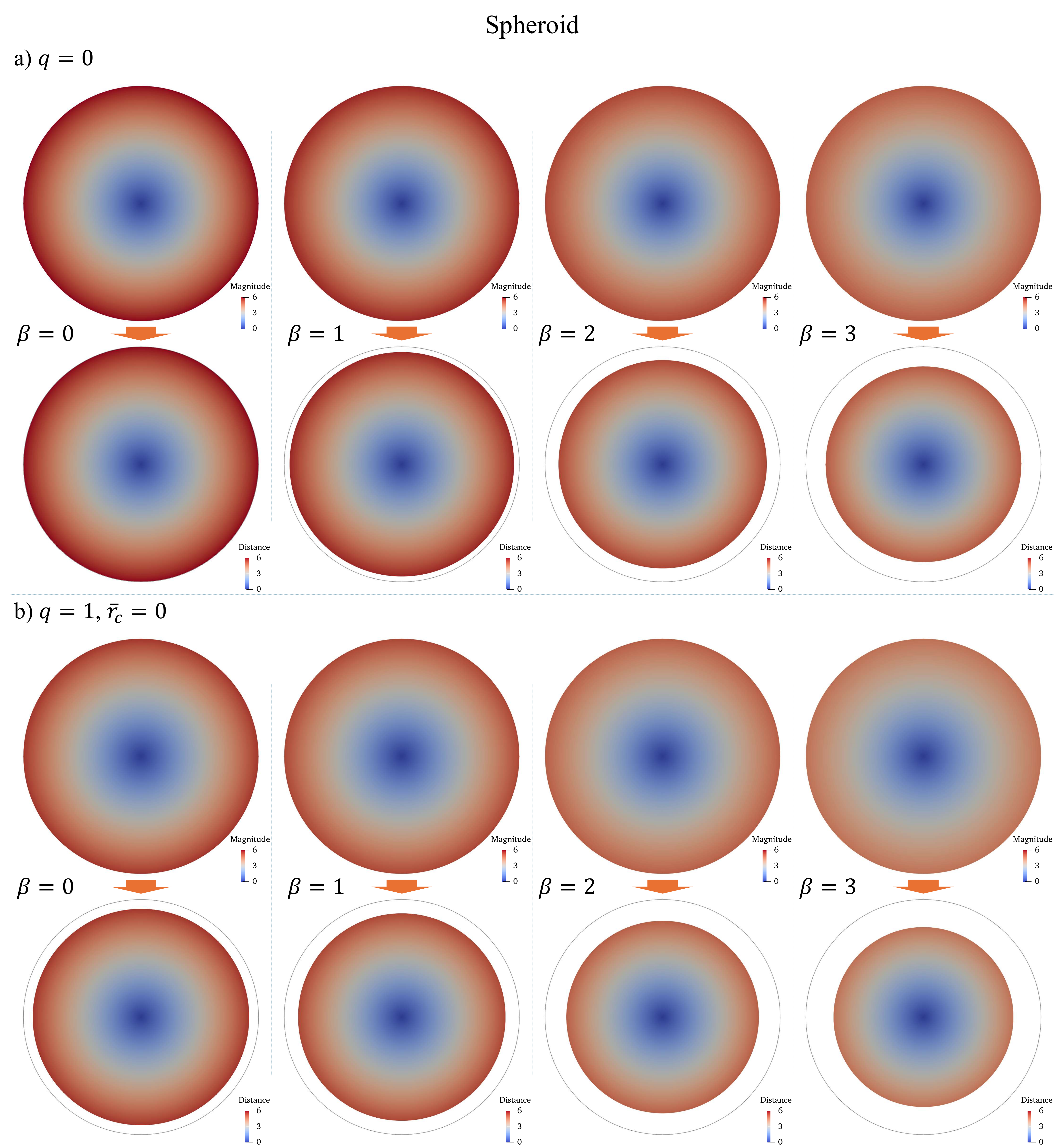}
    \caption{Solutions and corresponding actual domains for spheroids without (a) or with (b) a central disclination. The upper rows present the solutions on the reference domain, where the color indicates the magnitude of the vector field $\bm{x}(\bm{\bar{x}})$, i.e., $|\bm{x}|$. With the solution, we can relocate each point $\bm{\bar{x}}$ in the reference space to its corresponding position $\bm{x}$ in the actual space, which constitutes the lower rows. Here, the color indicates the distance between that point to the $z$-axis. The circles in the lower rows circumscribe the reference domains for comparison. $(\bar{r}_0, R_0)=(6, 10)$.}
    \label{si:fig:6}
\end{figure}

\begin{figure}
    \centering
    \includegraphics[width=\textwidth]{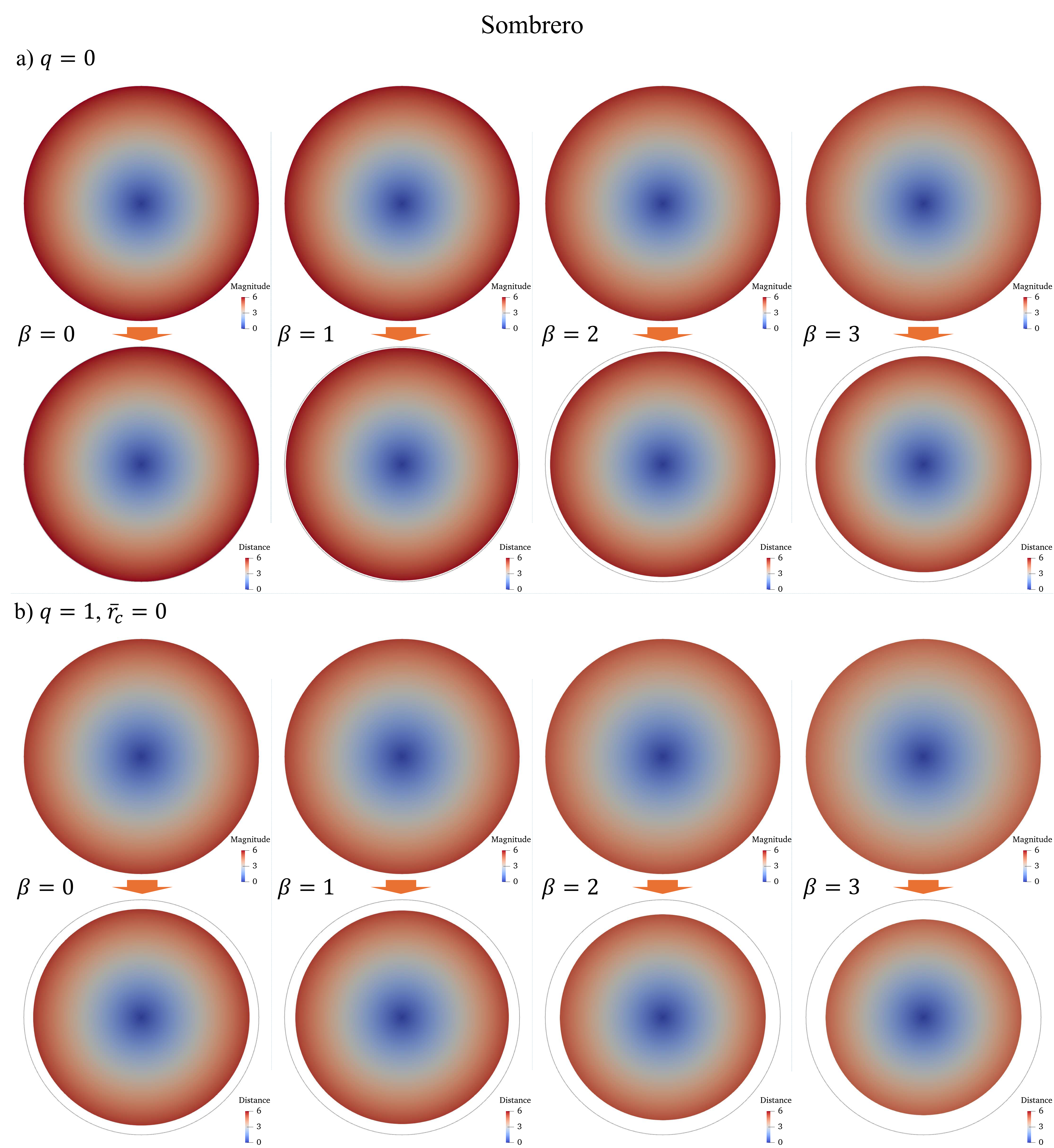}
    \caption{Solutions and corresponding actual domains for sombreros without (a) or with (b) a central disclination. The upper rows present the solutions on the reference domain, where the color indicates the magnitude of the vector field $\bm{x}(\bm{\bar{x}})$, i.e., $|\bm{x}|$. With the solution, we can relocate each point $\bm{\bar{x}}$ in the reference space to its corresponding position $\bm{x}$ in the actual space, which constitutes the lower rows. Here, the color indicates the distance between that point to the $z$-axis. The circles in the lower rows circumscribe the reference domains for comparison. $(\bar{r}_0, R_0)=(6, 10)$.}
    \label{si:fig:7}
\end{figure}

\begin{figure}
    \centering
    \includegraphics[width=0.8\textwidth]{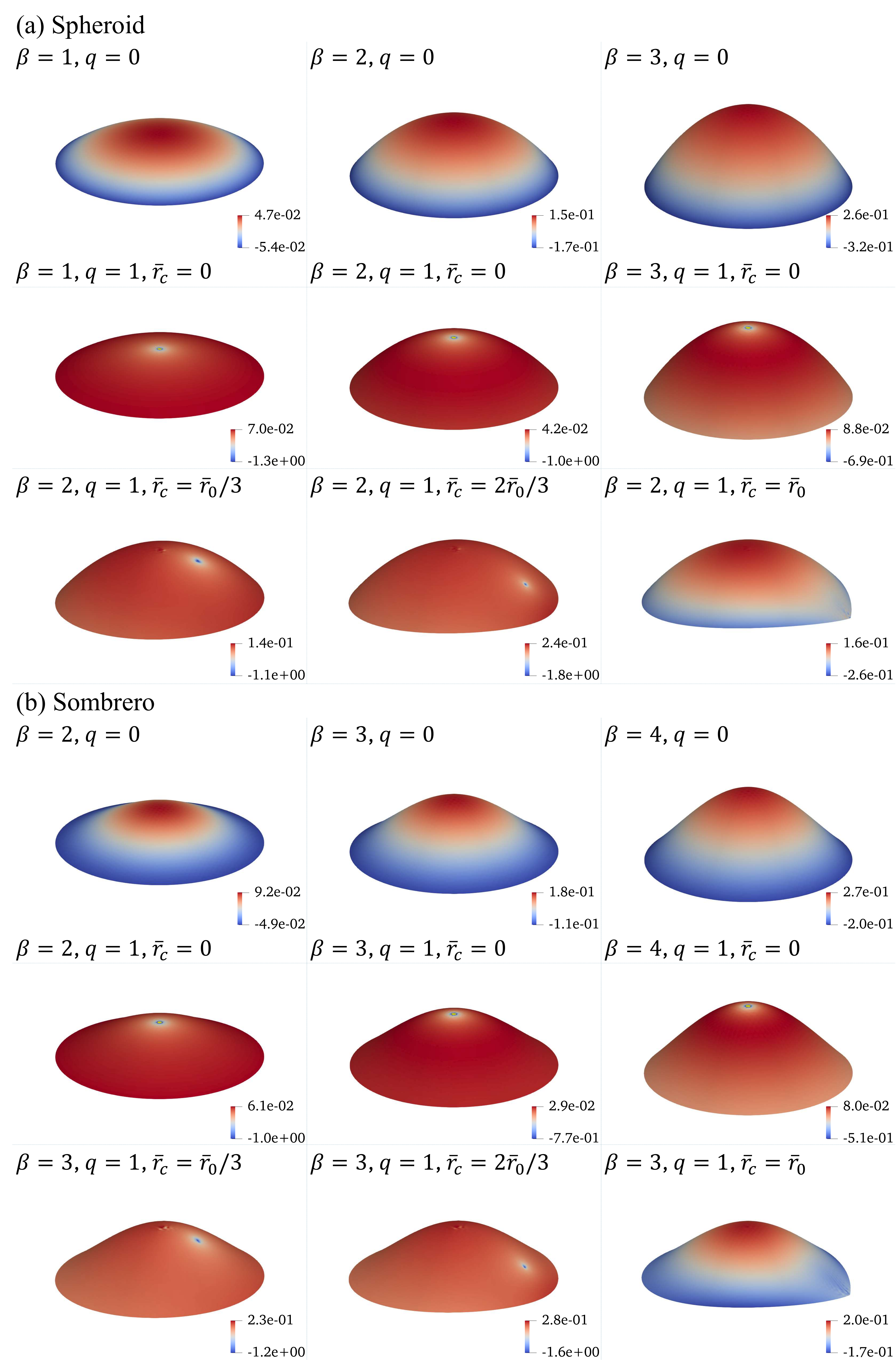}
    \caption{Stress maps in the actual space. For spheroid, $(\bar{r}_0, R_0)=(6, 10)$. For Sombrero, $(\bar{r}_0, R_0)=(6, 10)$}
    \label{si:fig:8}
\end{figure}

\begin{figure}
    \centering
    \includegraphics[width=0.9\textwidth]{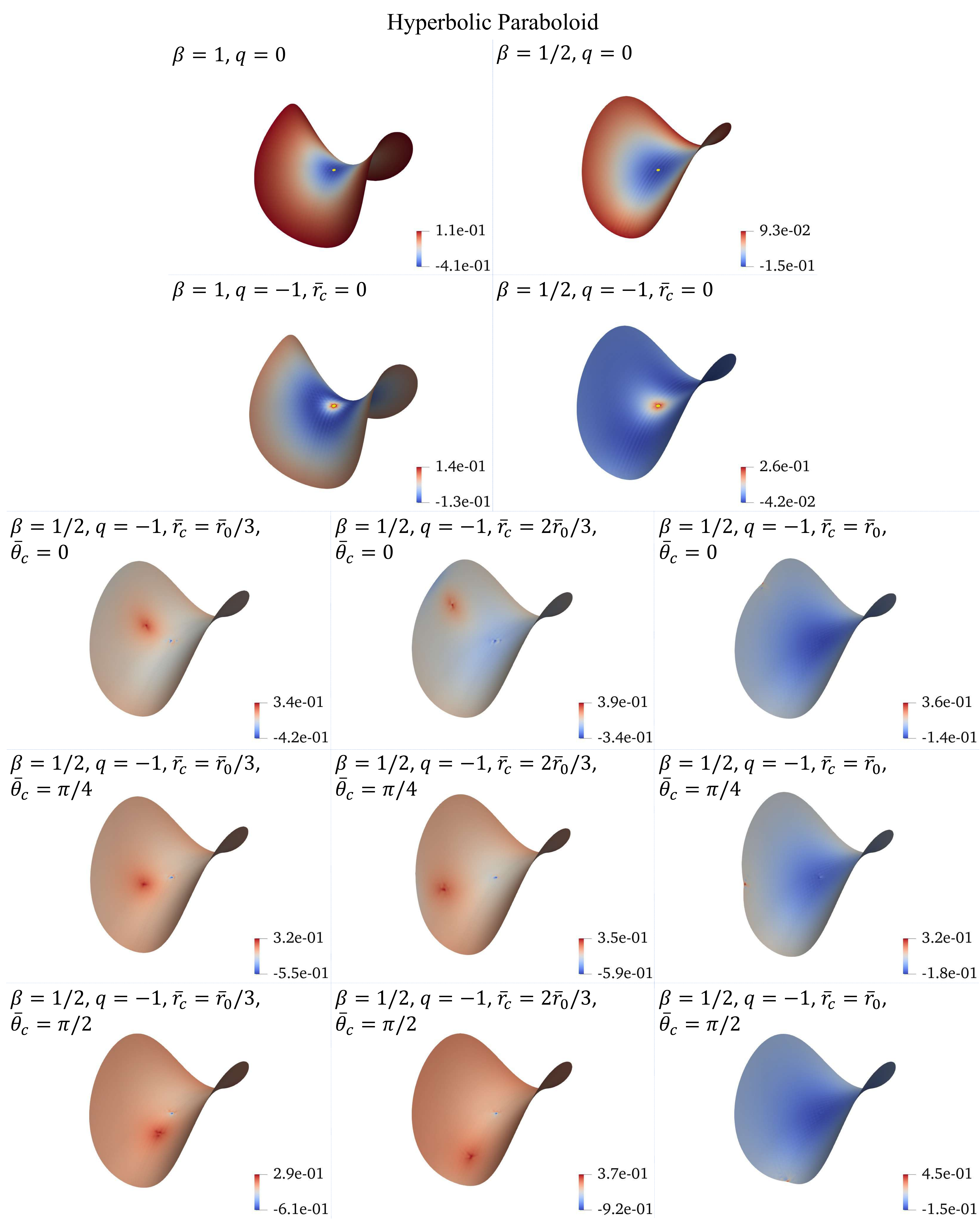}
    \caption{Stress maps in the actual space for hyperbolic paraboloid $(\bar{r}_0, R_0)=(96, 8)$}
    \label{si:fig:9}
\end{figure}

\begin{figure}
    \centering
    \includegraphics[width=\linewidth]{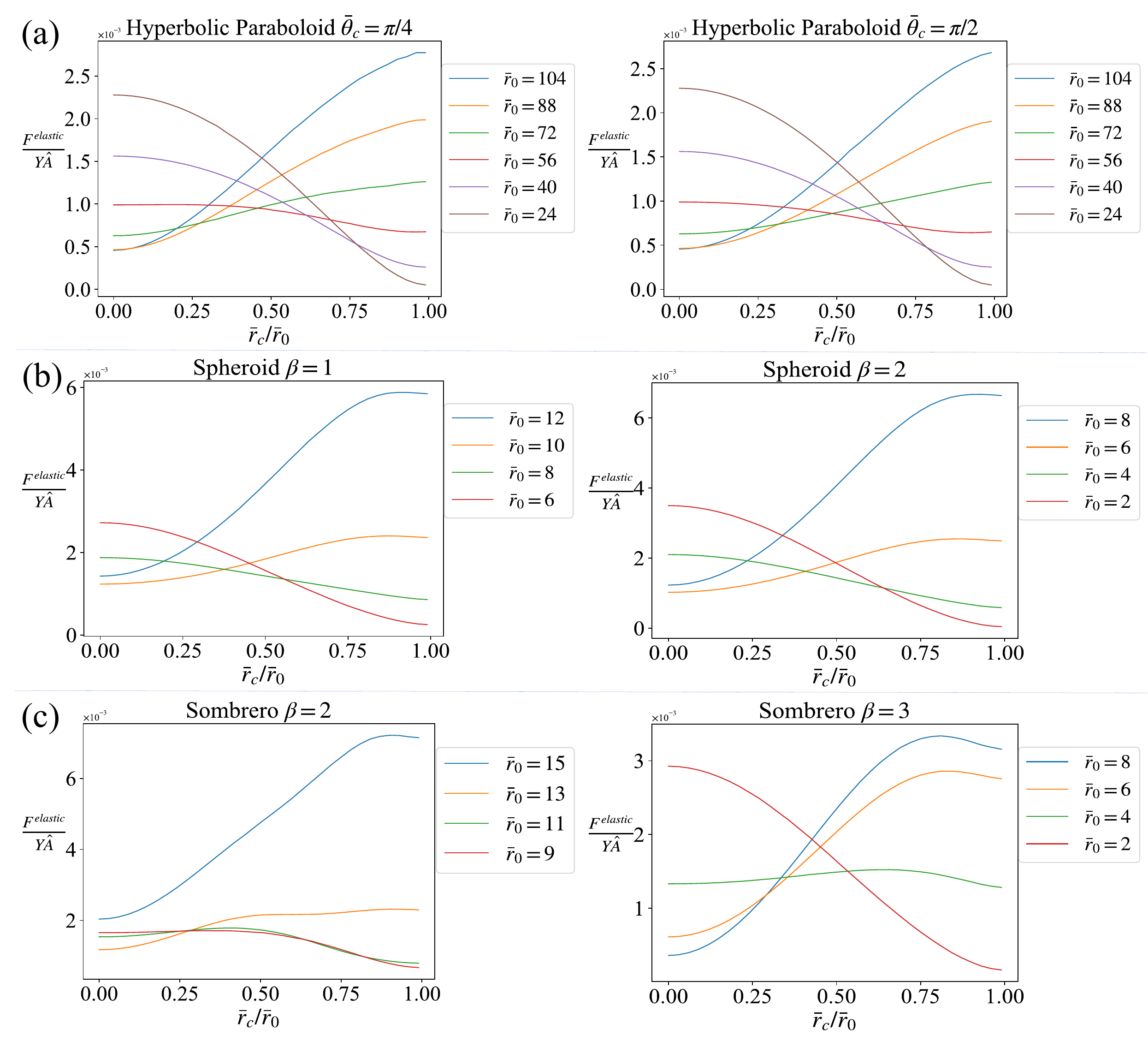}
    \caption{The energy as a function of distance between the disclination and the reference domain center for (a) hyperbolic paraboloid $\beta=1/2$, $R_0=8$, $\bar{\theta}_c = \pi/4, \pi/2$, (b) spheroid $\beta=1,2$, $R_0=10$ and (c) sombrero $\beta=2,3$, $R_0=10$. For spheroid and sombrero, the energy does not have angular dependence due to ratational symmetry.}
    \label{si:fig:10}
\end{figure}

\begin{figure}
    \centering
    \includegraphics{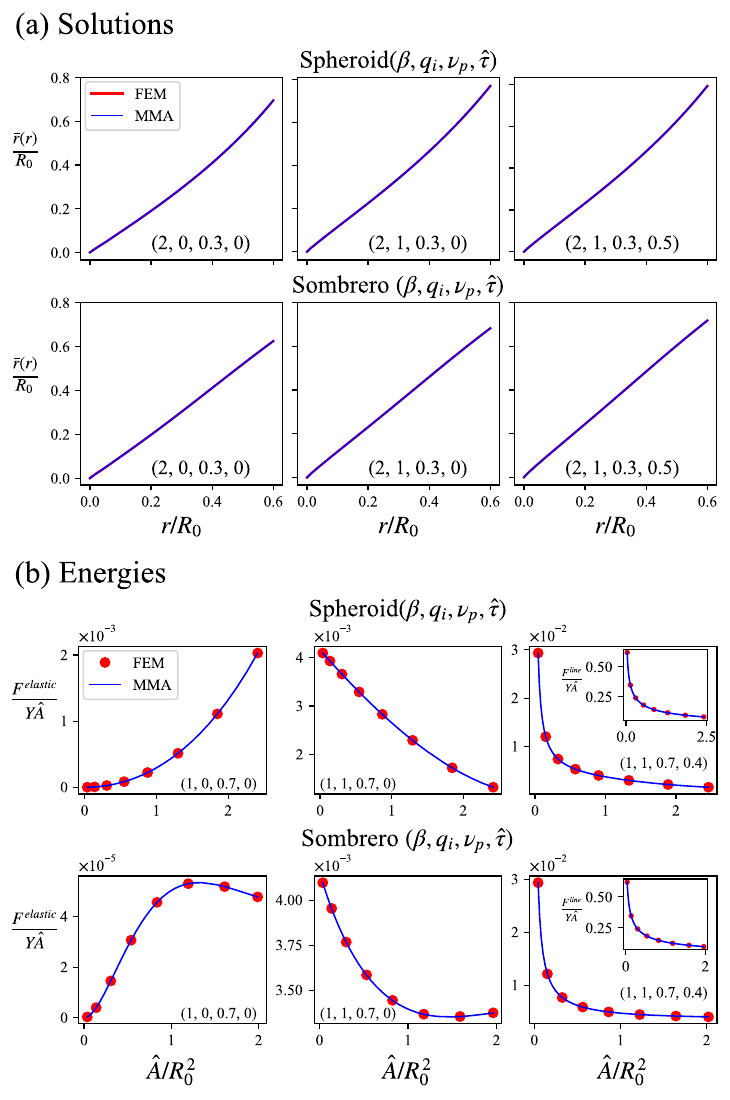}
    \caption{The reproducing of the results in Ref.~\cite{YinanMe2022}. The data in blue is obtained from Mathematica after simplifications from assuming the rotational symmetry. The data in red is obtained from the finite element method developed in the main text. (a) The plots of solutions. (b) The elastic energy $F^{elastic}$ versus the area of the reference domain $\hat{A}$. The inset is the line energy $F^{line}$ versus $\hat{A}$ when the line tension $\tau$ is nonzero.}
    \label{si:fig:2}
\end{figure}

\begin{figure}
    \centering
    \includegraphics[width=\linewidth]{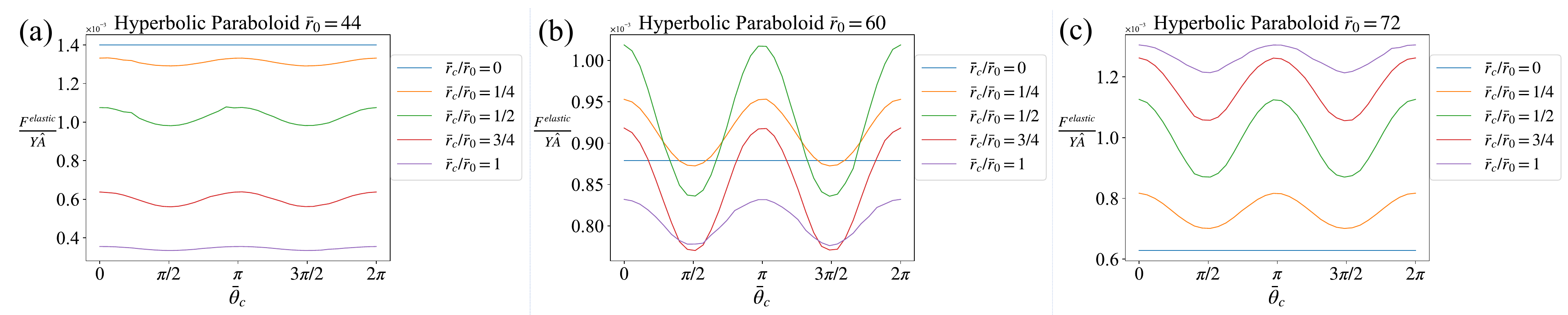}
    \caption{Energy as the disclination rotates at the given radius $\bar{r}_c$ on the circular reference domain with domain radius (a) $\bar{r}_0=44$, (b) $\bar{r}_0=60$ and (c) $\bar{r}_0=72$ on hyperbolic paraboloid for $\beta=1/2$, $R_0=8$. Numerical instabilities are noted for few points.}
    \label{si:fig:11}
\end{figure}

\begin{figure}
    \centering
    \includegraphics[width=\linewidth]{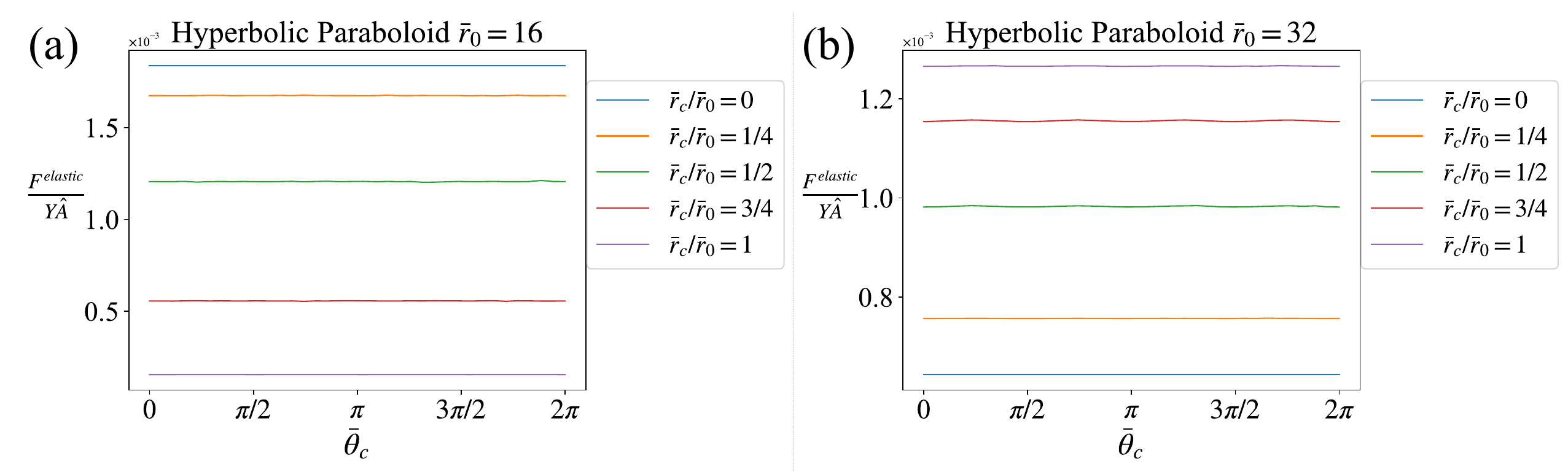}
    \caption{Energy as the disclination rotates at the given radius $\bar{r}_c$ on the circular reference domain with domain radius (a) $\bar{r}_0=16$ and (b) $\bar{r}_0=32$ on hyperbolic paraboloid for $\beta=1$, $R_0=8$.}
    \label{Fig:SI:energy:theta:beta:1}
\end{figure}

\begin{figure}
    \centering
    \includegraphics[width=\linewidth]{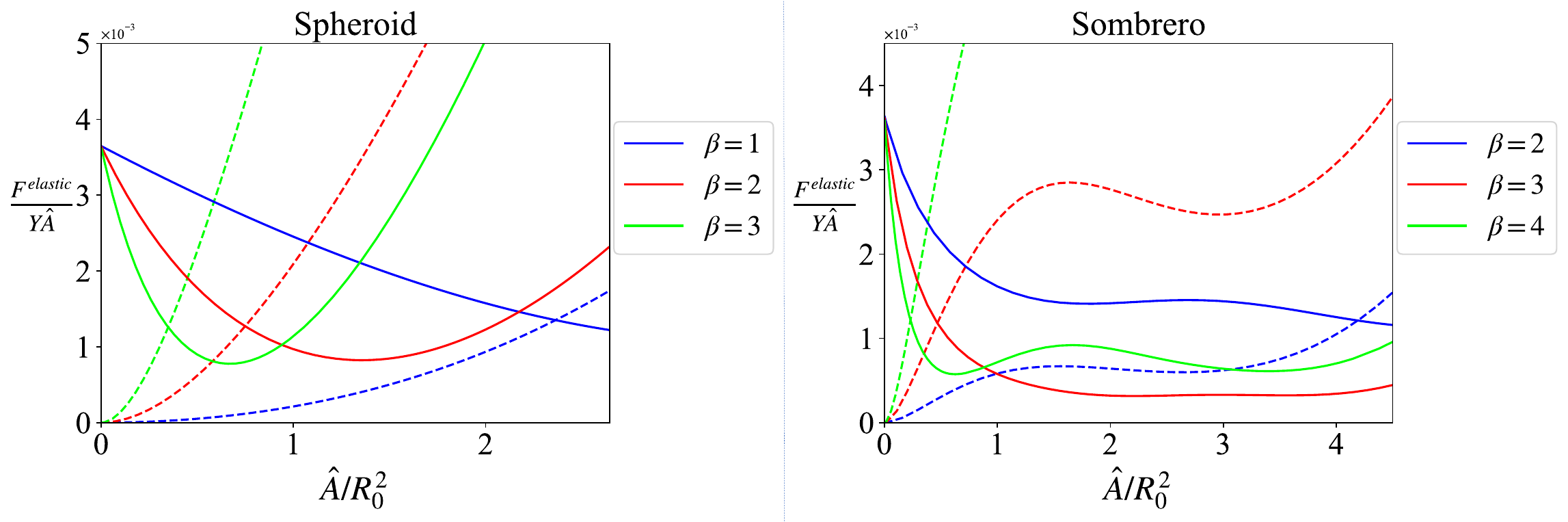}
    \caption{Energy vs.~the domain area in reference space for a square domain on a spheroid with $R_0=10$ for $\beta=1,2,3$ and on a sombrero with $R_0=10$ for $\beta=2,3,4$. The dashed lines represent the energy without a disclination, and the solid ones show energy with a centered 5-fold disclination. The intersection of the dashed and solid lines indicates the transition area where the formation of a discliantion at the center becomes energetically favorable.  The  intersection area values are listed in TABLE~\ref{table:si:1}.
    }
    \label{Fig:SI:13}
\end{figure}

\begin{figure}
    \centering
    \includegraphics[width=\linewidth]{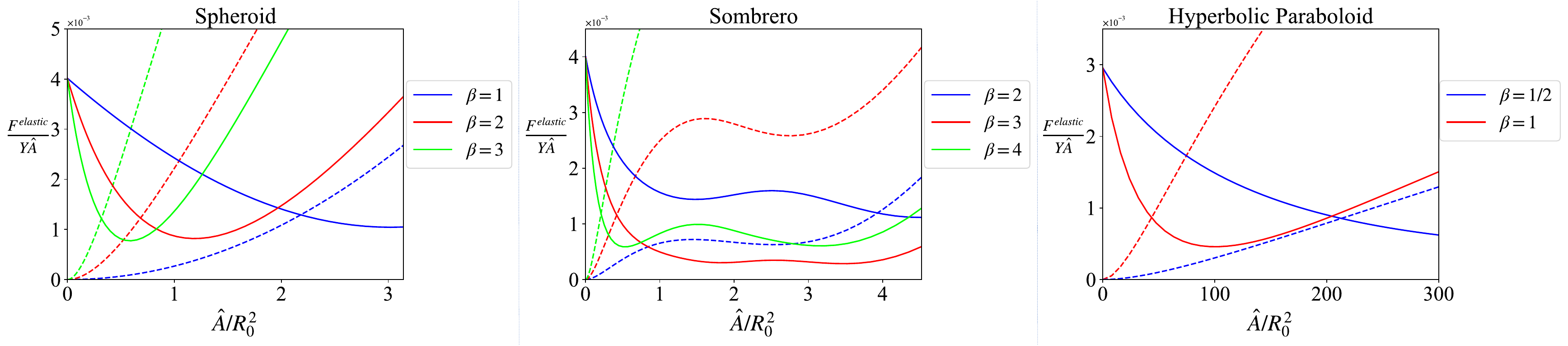}
    \caption{Energy vs.~the domain area in reference space for a circular domain on a spheroid with $R_0=10$ for $\beta=1,2,3$, on a sombrero with $R_0=10$ for $\beta=2,3,4$ on a hyperbolic paraboloid with $R_0=8$ for $\beta=1/2,1$. The dashed lines represent the energy without a disclination, and the solid ones show energy with a centered 5-fold or 7-fold disclination. The intersection of the dashed and solid lines indicates the transition area where the formation of a discliantion at the center becomes energetically favorable.  The  intersection area values are listed in TABLE~\ref{main:table:hyp_onset_list} and TABLE~\ref{table:si:1}.
    }
    \label{Fig:SI:14}
\end{figure}

\setcounter{figure}{0}
\renewcommand{\figurename}{TABLE}
\renewcommand{\thefigure}{S\arabic{figure}}

\begin{figure}
    \centering
    \includegraphics[width=\linewidth]{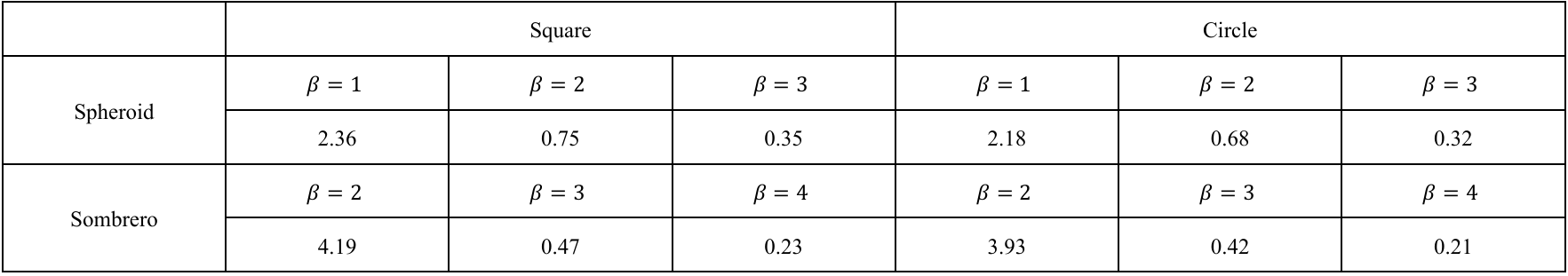}
    \caption{Transition normalized areas $A/R_0^2$ (see Figs.~\ref{Fig:SI:13}~and~\ref{Fig:SI:14}) beyond which a disclination becomes energetically favorable for both square and circular domains on a spheroid and a sombrero. The table highlights the difference between the transition areas of circular and square domains.
    }
    \label{table:si:1}
\end{figure}

\setcounter{figure}{14}
\renewcommand{\figurename}{FIG.}
\renewcommand{\thefigure}{S\arabic{figure}}

\end{document}